\documentclass[manuscript]{aastex}
\slugcomment{ApJ, in preparation}

\shorttitle{Substructure in M33}
\shortauthors{McConnachie et al.}

\begin{document}

\title{The photometric properties of a vast stellar substructure in the
outskirts of M33}

\author{Alan W. McConnachie$^1$\email{alan.mcconnachie@nrc-cnrc.gc.ca}} 
\author{Annette M. N. Ferguson$^2$, Michael J. Irwin$^3$, John Dubinski$^4$,
  Lawrence M. Widrow$^5$, Aaron Dotter$^6$, Rodrigo Ibata$^7$, Geraint F.
Lewis$^8$} 
\affil{
$^1$NRC Herzberg Institute of Astrophysics, 5071
West Saanich Road, Victoria, B.C., V9E 2E7, Canada
$^2$Institute for Astronomy, University of Edinburgh, Royal Observatory, Blackford Hill, Edinburgh EH9 3HJ, UK
$^3$Institute of Astronomy, University of Cambridge, Madingley Road, Cambridge CB3 0HA, UK.
$^4$Department of Astronomy \& Astrophysics, University of Toronto, 50 St. George Street, Toronto, Ontario, Canada M5S 3H4
$^5$Department of Physics, Engineering Physics, and Astronomy Queen's University, Kingston, Ontario, Canada K7L 3N6 
$^6$Department of Physics and Astronomy, University of Victoria, 3800
Finnerty Road, Victoria, British Columbia, Canada V8P 5C2
$^7$Observatoire de Strasbourg, 11, rue de l'Universit{\'e}, F-67000 Strasbourg, France
$^8$Sydney Institute  for Astronomy, School  of Physics, A29, University of  Sydney, NSW 2006, Australia}

\clearpage
\newpage

\begin{abstract}

  We have surveyed approximately 40 square degrees surrounding M33
  with CFHT MegaCam/MegaPrime in the $g$ and $i$ filters out to a
  maximum projected radius from this galaxy of 50\,kpc, as part of the
  {\it Pan-Andromeda Archaeological Survey (PAndAS)}. Our observations
  are deep enough to resolve the top $\sim 4$ magnitudes of the red
  giant branch population in this galaxy. We have previously shown
  that the disk of M33 is surrounded by a large, irregular,
  low-surface brightness substructure. Here, we quantify the stellar
  populations and structure of this feature using the PAndAS data. We
  show that the stellar populations of this feature are consistent
  with an old population with $<[Fe/H]> \sim -1.6$\,dex and an
  interquartile range in metallicity of $\sim 0.5$\,dex. We construct a
  surface brightness map of M33 that traces this feature to $\mu_V
  \simeq 33$\,mags\,arcsec$^{-2}$. At these low surface brightness
  levels, the structure extends to projected radii of $\sim 40$\,kpc
  from the center of M33 in both the north-west and south-east
  quadrants of the galaxy. Overall, the structure has an ``S-shaped''
  appearance that broadly aligns with the orientation of the HI disk
  warp. We calculate a lower limit to the integrated luminosity of the
  structure of $-12.7 \pm 0.5$\,mags, comparable to a bright dwarf
  galaxy such as Fornax or Andromeda~II and slightly less than $1\%$
  of the total luminosity of M33. Further, we show that there is
  tentative evidence for a distortion in the distribution of young
  stars near the edge of the HI disk that occurs at similar azimuth to
  the warp in HI. The data also hint at a low-level, extended stellar
  component at larger radius that may be a M33 halo component. We
  revisit studies of M33 and its stellar populations in light of these
  new results, and we discuss possible formation scenarios for the
  vast stellar structure. Our favored model is that of the tidal
  disruption of M33 in its orbit around M31.

\end{abstract}


\clearpage
\newpage

\section{Introduction}

\cite{rogstad1976} noted that they had succeeded in ``turning three
puzzles into one enigma'' with their analysis of the Local Group
spiral galaxy M33 (the Triangulum Galaxy). These authors postulated
that a severe warp in its HI disk could explain simultaneously three
phenomena observed in their 21cm dataset; the ``wings'' of that galaxy, its
steep radial profile, and the ``so-called weak component''. Subsequent
radio observations have shown their postulate to be correct
(\citealt{corbelli1997,putman2009}), but exactly why this galaxy
formed a warp in the first place, and why no optical counterpart has been
observed, have remained enigmatic since this pioneering study.

M33 ($1^h 33^m 51^s, 30^\circ 39^\prime 36^{\prime\prime}$) is the
third brightest spiral galaxy in the Local Group, with a luminosity of
$M_V = -18.9$. It is a late-type spiral, Sc II-III, and shows no clear
evidence of any bulge component
(\citealt{bothun1992,minniti1993,mclean1996}). Its rotation curve
indicates a total mass of $>5 \times 10^{10}$\,$M_\odot$ within 16kpc
(the radius of the last measured point; \citealt{corbelli2000}),
approximately one-tenth that of its giant neighbor, M31. There is
considerable disagreement in the literature regarding the distance of
M33, with estimates ranging from $\sim 800$\,kpc to $>900$\,kpc (e.g.,
\citealt{galleti2004,ciardullo2004,tiede2004,mcconnachie2004a,mcconnachie2005a,brunthaler2005,bonanos2006,sarajedini2006,u2009}
and references therein). However, given the distance of M31 ($D_{M31}
= 785 \pm 25$\,kpc; \citealt{mcconnachie2005a}) and the separation of
M31 and M33 on the plane of the sky ($\sim 15$\,degrees), it is
reasonable to consider M33 as the most luminous satellite of M31.

A deep survey of M33 with the 2.5 meter Isaac Newton Telescope Wide
Field Camera (INT WFC), that resolves individual stars in this galaxy
down to $V = 24.5$ and $i = 23.5$ (Vega mags) with $S/N \sim 5$, shows a
relatively ``pristine'' galaxy, with no obvious distortions or
substructures over the area surveyed (\citealt{ferguson2007}). This is
in stark contrast to results for M31, where an identical survey over
an initial survey area of $\sim 25$\,square degrees
(\citealt{ferguson2002}; later extended to $\sim 40$\,square degrees;
\citealt{irwin2005}) revealed copious substructures. The most
prominent of these was a giant stellar stream (\citealt{ibata2001a})
that reached to the edge of the survey area, as well as various other
streams, overdensities and new dwarf galaxies (e.g., \citealt{zucker2004a,
  zucker2004b, mcconnachie2003, mcconnachie2004b,
  irwin2008}). That M33 does not display the same richness of
structures as M31 may in whole or in part be due to its
significantly lower mass (e.g., \citealt{purcell2007}).

A subsequent survey of M31 and M33 by \cite{ibata2007}, using the
1\,degree field-of-view MegaCam/MegaPrime camera on the 3.6 meter
Canada-France-Hawaii telescope (CFHT), identified stars to the deeper
magnitude limits of $g \simeq 25.5$, $i \simeq 24.5$ (AB mags) at a $S/N =
10$. They contiguously map the entire south-east quadrant of M31 to a
radius of 10\,degrees ($\sim 150$\,kpc), and extend the survey area
along the southern minor axis of M31 to connect with M33. They trace
the radial surface brightness profile of M31 as it declines in the
south-east quadrant to large radius, at which point the profile starts
to increase again (see their Figure~38). \cite{ibata2007} attribute
this rise to a direct detection of the M33 stellar halo.

Building upon these previous surveys, the {\it Pan-Andromeda
Archaeological Survey} (PAndAS; \citealt{mcconnachie2009b}), a Large
Program on the CFHT, is contiguously mapping the entire halo of M31
out to a maximum projected radius of 150\,kpc, and the surroundings of
M33 out to a maximum projected radius of 50\,kpc.  One of the most
striking discoveries presented in \cite{mcconnachie2009b} from the
first year of the survey is the presence of a very large, low surface
brightness, stellar substructure surrounding M33, and which extends
out to several times the radius of the classical disk of M33. In this
contribution, we examine in more detail the properties of this vast
structure, as observed with the CFHT/MegaPrime data.

The format of this paper is as follows. In Section 2, we summarize the
design on the PAndAS survey and overview the data as it
relates to this study. In Section 3, we analyze and quantify the
structure and stellar populations of the M33 substructure. In Section
4, we discuss our results in the context of the evolutionary history
of M33 and previous studies of its disk and halo populations. Section
5 summarizes. Throughout this paper, we adopt a distance modulus for
M33 of $\left( m - M \right)_0 = 24.54 \pm 0.06$ ($809 \pm 24$\,kpc;
\citealt{mcconnachie2004a,mcconnachie2005a}).

\section{Preliminaries}

\subsection{Survey strategy}

PAndAS is a Large Program on the 3.6 meter Canada-France Hawaii
Telescope which uses the $0.96 \times 0.94$ degree field of view
MegaPrime wide field camera to obtain $g, i$ imaging of M31, M33 and
their environs. This instrument consists of a mosaic of 36 $2048
\times 4612$ pixel CCDs with a pixel scale of 0.187
arcsec\,pixels$^{-1}$ at the center of the detector. Data acquisition
for PAndAS started in the 2008B observing semester (S08B) and will
continue until S10B, with the aim of contiguously imaging M31 and M33
out to projected radii of $\sim 150$\,kpc and $\sim 50$\,kpc,
respectively, representing a total area of $> 300$\,square degrees.
This survey builds upon earlier CFHT/MegaPrime imaging of M31 and M33
from PI programs by R. Ibata (\citealt{martin2006,ibata2007}) and
A. McConnachie (\citealt{mcconnachie2008b}). First results from
PAndAS, showing the full extant survey area as at the end of S08B, are
given in \cite{mcconnachie2009b}.

Figure~1 shows the location of 40 MegaPrime fields centered on M33 in
a tangent plane projection, where open squares represent
CFHT/MegaPrime observations taken in S08B, and hatched fields show
earlier fields presented in \cite{ibata2007}.  The solid angle subtended by
these fields is equivalent to $\sim 7800$\,kpc$^2$ at the distance of
M33. The inner ellipse represents the ``edge'' of the optical disk of
M33 ($70.8 \times 41.7$\,arcmins diameter\footnote{Values taken from the
NASA Extragalactic Database}, the radius at which the disk surface
brightness reaches $\mu_B \simeq 25$\,mags\,arcsec$^{-2}$). The major
and minor axes are shown, and the dashed outer circle corresponding to
a maximum projected radius of 50\,kpc at the adopted distance of M33.

We adopt a similar observational strategy to the earlier PI programs,
whereby we expose for 1350\,seconds in $g$ and $i$, split into $3$
dithered sub-exposures of $450$\,seconds. All of our observations were
taken in generally excellent seeing conditions, with a median $FWHM$
in the new data of better than $0.65$\,arcsecs. This is generally
sufficient to reach $g \simeq 25.5$ and $i \simeq 24.5$ (AB mags) with $S/N =
10$. In some cases, more than three exposures were taken (at the
discretion of CFHT staff to ensure the requested observing conditions
were met), and in these cases the viable images were included in the
stacking procedure, weighted according to noise/seeing. 

\subsection{Data processing and calibration}

The CFHT/MegaPrime data were pre-processed by CFHT staff using the
{\it Elixir} pipeline, which accomplishes the usual bias, flat, and
fringe corrections and also determines the photometric zero point of
the observations. These images were then processed using a version of
the CASU photometry pipeline (\citealt{irwin2001}) adapted for
CFHT/MegaPrime observations. The pipeline includes re-registration,
stacking, catalogue generation and object morphological
classification, and creates band-merged $g,i$ products for use in the
subsequent analysis.  The CFHT $g$ and $i$ magnitudes are de-reddened
on a source-by-source basis using the \cite{schlegel1998} IRAS maps,
such that $g_0 = g - 3.793 E(B - V)$ and $i_0 = i - 2.086 E(B - V)$,
where $g_0$ and $i_0$ are the de-reddened
magnitudes. \cite{regnault2009} discuss many issues relating to the
calibration of CFHT/MegaPrime photometry, and the data presented here are
accurate at the level of $\pm 5\%$ peak-to-peak (judging from overlap
regions between fields). This is sufficiently
accurate for the present analysis, and a future contribution will
discuss in more detail the calibration of the dataset.

The $i-$band photometry of the hatched fields in Figure~1 (observed in
S05B) was taken with a different filter than the other fields
(observed in S08B) due to the filter being replaced by CFHT in
2007. The correction between the two generations of photometry is
expected to be small, and we derive it by searching for common objects
in areas of overlap between the earlier imaging and the new data. To
ensure a high quality fit, we use only those objects robustly
identified as stellar in all 4 filters (the old and new $i$ filter,
and the two generations of $g-$band data), which do not lie near the
edges of CCDs, and which have photometric errors less than 0.1\,mags
in all filters. Figure~2 shows the difference in the $i-$band
photometry as a function of color (where $i_1$ refers to the new
$i-$filter and $i_2$ refers to the old $i-$filter). The two filters
have similar throughput at a color of $(g - i) \simeq 1.6$ and there
is a slight gradient over the color range of interest
($\sim0.07$\,mags between $(g-i_2) = 0.5$ and $(g-i_2) = 2.0$). The
red points and error bars show the binned data (50 points per bin) and
the blue line shows the best linear fit to the data, described by

\begin{equation}
i_2 - i_1 = -0.045 (g - i_2) + 0.073~.
\end{equation}

\noindent We use Equation 1 to transform all $i-$band photometry taken
with the previous $i-$filter onto the same scale as the new $i-$band
data.

Our final photometric uncertainties as a function of magnitude are
shown in Figure~3 for each filter. Data for the central field, where
crowding is very significant, is not shown. The distinct sequences in
Figure~3 are due to fields taken in different observing
conditions. Generally, the mean uncertainty is smaller than 0.2\,mags
for $g \le 25.5$ and $i \le 24.5$.

\subsection{Stellar isochrones}

Figure 4 shows a comparison between the throughput of the CFHT and
SDSS filter systems. While the systems are fairly similar in $r$, in
the $g$ and $i$ filters there are some differences. In particular, the
CFHT $g$ filter is shifted to the red in comparison to the SDSS
equivalent, and the CFHT $i$ filter is broader than the SDSS equivalent by $\sim 250\AA$ towards the red end of
the spectrum. Since we have well-determined atmospheric transmissions
for Mauna Kea and for the overall throughput of CFHT, and that these
are different to their SDSS counterparts, we decided to construct
isochrones in the CFHT photometric system. To achieve this, isochrones
from the Dartmouth Stellar Evolution Database
(\citealt{dotter2007,dotter2008}) were transformed to the CFHT system
using filter response functions for the five broadband filters,
combined with the throughput of the telescope optics, the quantum
efficiency of the MegaPrime CCD, and the atmospheric transmission at
Mauna
Kea\footnote{\url{http://www.cfht.hawaii.edu/Instruments/Imaging/MegaPrime/specsinformation.html.}}. The
magnitudes were normalized to Vega using the flux-calibrated spectrum
provided by \cite{bohlin2007} and then adjusted to the CFHT AB
magnitude system by applying the appropriate offsets\footnote{\url{http://www.cfht.hawaii.edu/Instruments/Imaging/MegaPrime/specsinformation.html}}.
Bolometric corrections were then obtained using the grids of synthetic
spectra described in Section~4 of \cite{dotter2008}. A grid of
isochrones was created for $-2.5 \leq$ [Fe/H] $\leq 0$\,dex at
[$\alpha$/Fe] = 0, +0.2, and +0.4\,dex and [Fe/H] = +0.3 and +0.5\,dex
at [$\alpha$/Fe] = 0 and +0.2\,dex. 

Figure~5 shows a comparison between the CFHT isochrones (dashed lines)
and globular cluster fiducial sequences for M92, M3 and M71 (solid
lines, left to right) from \cite{clem2008}.  The assumed apparent
distance moduli in $g$ and adopted E(B-V) are (14.64, 0.04), (15.05,
0.03) and (13.79, 0.26) for M92, M3 and M71, respectively. For each
cluster, we compare to 14Gyr isochrones with appropriate [Fe/H] and
[$\alpha$/Fe] values, as given in the key in Figure~5. Over the
displayed magnitude range, the isochrones provide excellent matches to
the globular cluster sequences, implying that these isochrones are a
good reference system for which to compare to real stellar
populations. We note, however, that for this particular study we are
concerned only with the RGB phase of stellar evolution. For reference,
at [Fe/H] $= -1.5$\,dex for a 12Gyr population, the isochrones shift
redward by 0.05\,mags in going from [$\alpha$/Fe] = 0 to [$\alpha$/Fe]
= +0.2. Note that the Dartmouth isochrone set used here are generally
redder at a given metallicty on the RGB than the Padova isochrones
(\citealt{girardi2004}) used by \cite{ibata2007}, leading to maximum
difference in RGB metallicity for individual stars of a few tenths of
a dex (see the comparison in \citealt{dotter2007}).

\section{Photometric analysis of the substructure surrounding M33}

\subsection{Global color-magnitude diagrams}

Figure~6 shows the overall $g_o$ and $i_o$ color-magnitude diagrams
(CMDs; left and right panels, respectively) for all stellar sources in
the M33 survey area (excluding the central field). Only stars robustly
identified as stellar in both filters are shown, corresponding to $>
650\,000$ stars. The CMDs are plotted as Hess images with $0.025
\times 0.025$\,mag pixels and displayed with logarithmic
scaling. Overlaid on each CMD are isochrones for a 12 Gyr,
[$\alpha/$Fe] $=0.0$, stellar population shifted to the distance
modulus of M33, with $[Fe/H] = -2.5, -2, -1.5, -1$ and $-0.5$\,dex.

Several features are obvious in Figure~6, not all of which represent
components of M33 and its surroundings. The near-vertical sequences in
both CMDs at $(g - i)_o \simeq 0.1$ are the main-sequence turn-off
(MSTO) populations in the Galactic halo along our line of sight to
M33. Two distinct main sequence features are visible in each CMD,
indicating the presence of considerable (sub)structure in this region
of the Milky Way halo (e.g., see
\citealt{rochapinto2004,martin2007}). In addition to the Galactic
halo, a contribution from the Galactic disk is also visible in
Figure~6: the majority of stars redder than $(g-i)_o > 2$ at all
magnitudes represent the locus of foreground dwarfs in the disk of
the Milky Way, and the density distribution of this population
increases to the north of the survey. Finally, compact galaxies at
typical redshifts of $z \sim 0.5$ that are misidentified as stars
become a significant source of contamination at faint magnitudes
($g_0, i_0 > 24$), and tend to have predominantly bluer colors.

The stellar isochrones in Figure~6 show the locus of the CMD that
contains red giant branch (RGB) stars in M33. Approximately 200\,000
stellar objects are bounded by these isochrones (excluding the central
field). At faint magnitudes, contamination of this locus increases due
to the galaxies misidentified as stars; at redder colors (typically
more metal-rich RGB stars at a given age, or older RGB stars at a
given metallicity) contamination of this locus increases due to the
foreground Galactic disk stars. If present, asymptotic giant branch
(AGB) stars, typically representative of an intermediate-age stellar
population, may occupy a similar locus to the RGB, and can extend to
magnitudes significantly brighter than the tip of the RGB (e.g.,
Table~7 of \citealt{rejkuba2006}). Young main sequence and blue-loop
stars are also visible in Figure~6 as bright sources with $(g-i)_o <
0$.

\subsection{Stellar population maps}

\subsubsection{Contaminants}

Prior to examining the spatial distribution of the stellar populations
in M33, it is important to understand the distribution of sources that
could act as contamination and influence our results. To this end,
Figure~7 shows tangent plane projections of the density distribution
of the various types of sources that can act as contamination in our
study of the M33 stellar populations. These maps use a pixel size of
$1.2 \times 1.2$\,arcmins, are unsmoothed and are displayed with
linear scaling. The left panel shows stellar sources in the Galactic
halo, here traced by stars within the color-magnitude cut $19.0 < i_0
< 22.0$, $0.1 < (g - i)_0 < 0.6$ (halo turnoff stars). In the central parts of M33, this
color cut picks up some blue loop stars in the disk of the
galaxy. The central panel shows stellar sources in the Galactic disk,
here traced by stars within the color-magnitude cut $17.0 < i_0 <
20.0$, $1.5 < (g - i)_0 < 3.0$ (red disk dwarfs). In the central parts of M33, this
color cut picks up some bright AGB and red supergiant stars in the
disk of the galaxy. The right panel shows extended sources (galaxies)
identified by our star-galaxy classification algorithm, within the
luminosity range $17.0 < i_0 < 23.5$. Any faint, compact, galaxies
that have erroneously been classified as stars are expected to have a
broadly similar distribution to the overall galaxy population
(extended sources). In the
main body of M33, the high degree of crowding causes some blended stellar
sources to be classified as extended sources; in the very central
regions of M33, crowding prevents sources being identified at fainter
magnitudes, creating an apparent hole in the distribution.

The Galactic halo and disk stars are relatively smoothly distributed
in Figure~7, insofar as there are no obvious large scale
substructures. The Galactic disk is located towards the north of the
survey but the gradient is not so extreme that it shows up with the
linear scaling of Figure~7 (M33 has Galactic coordinates $l =
133.6^\circ, b = -31.3^\circ$). The galaxy distribution is distinct to
the stars, and shows clear evidence of clustering.

\subsubsection{The red giant branch population}

Figure~8 shows a tangent plane projection of the density distribution
of all candidate RGB stars with $i_0 \le 23.5$. This limit has been
chosen to ensure a very high completeness level in the data and small
photometric uncertainties. These stars lie in the color-magnitude
locus bounded by 12\,Gyr, $[\alpha/Fe] = 0.0$, isochrones, with
metallicities between $-2.5 \le [Fe/H] \le -0.2$\,dex, shifted to the
distance modulus of M33. This map was created with $1.2 \times
1.2$\,arcmins pixels, then repeatedly smoothed 3 times using a linear
filter with a width of 5 pixels, and displayed with logarithmic
scaling.

In Figure~8, the concentration at $(\xi, \eta) \approx (-2.1^\circ,
2.6^\circ)$ is due to the globular cluster system of the background
elliptical galaxy NGC\,507. Due to the excellent seeing of the
observations, these background globular clusters appear as individual point sources with colors
similar to blue RGB stars. The slight north-to-south gradient visible
over the image is due to contamination from the Galactic disk in the
redder part of the RGB. The newly discovered M31/M33 satellite,
Andromeda~XXII (\citealt{martin2009}), is barely visible in this plot
at $(\xi, \eta) \approx (-1.4^\circ, -2.6^\circ)$. The disk of M33 is
visible, and clearly extends further than the radius marked by $\mu_B \simeq
25$\,mags\,arcsec$^{-2}$.

Of particular interest to this contribution is the low-level,
irregular distribution of RGB stars surrounding the disk of M33,
discovered by \cite{mcconnachie2009b}. Extensions stretching toward
the north-west and south-east, broadly aligned with the direction to
M31, are visible in Figure~8; the component in the north-west appears
to reach a maximum projected radius from M33 of $\sim 3$ degrees
(42\,kpc), 3 times further out than the main disk of the galaxy. In
the south-east, the extension appears to reach to a projected radius
of roughly 2\,degrees (28\,kpc).

\subsection{Color-magnitude diagrams and photometric metallicity}

Red outlines in Figure~8 show regions selected to
probe the stellar populations in the outer regions of M33. The
red elliptical annulus and the rectangular areas labeled H1, H2, S1, S2
and S3 probe the outer part of the M33 disk, two putative halo fields,
two fields on the north-western extent of the substructure and one
field on the southern substructure, respectively. Each of these has an
area of 0.35 square degrees. H1 and H2 are placed at similar
distances from M33 as the substructure fields and probably contain
broadly similar numbers of halo stars. The disk annulus has a mean semi-major
axis of $a = 49.6$\,arcmins ($11.7$\,kpc) and a width on the major
axis of 7.1\,arcmins (1.7\,kpc). The larger rectangular areas, labeled A
and B, are one degree wide strips in Galactic latitude (centered at $l
= 135.7^\circ$ and $l = 130.9^\circ$, respectively), that probe large
areas of the survey on the east and west side of M33, and which will
be used to explore the variation of foreground stellar
populations with Galactic latitude. 

Figure~9 shows the CMDs for the disk annulus field, H1, H2 (top row,
left to right, respectively), S1, S2 and S3 (middle row, left to
right, respectively), and areas A and B (bottom row, left and right,
respectively). Isochrones corresponding to [Fe/H] $= -0.5, -1.0, -1.5,
-2.0, -2.5$\,dex are overlaid on area B for reference. The disk field
appears to have the largest mix of stellar populations; in addition to
the foreground populations, a broad RGB is visible, indicating a
significant range in age and/or metallicity. A population of blue, $(g
- i)_o < 0$, stars is also visible, and a luminous AGB population may
also be present but is difficult to discern from the bright RGB and
Galactic disk foreground population. This field stands in contrast to
the two putative halo fields. In both H1 and H2, the only stellar
populations that are present in any significant number are the
foreground populations and faint, bluish `stars' that could all be
misidentified galaxies.  However, H2 in particular shows some evidence
of a very weak RGB. Thus a low level halo population in M33 cannot be
ruled out, and may vary spatially. This will be investigated in more
detail in a future contribution; here our focus is towards the large
stellar substructure.

In Figure~9, it is striking that the CMDs of S1, S2 and S3 are all
similar, and distinct to the disk annulus and putative halo fields. In
particular, the similarity of the S3 field with S1 and S2, despite
being located on the opposite side of the M33 disk, is strongly
suggestive of a connection between the substructure in the south-east
and that in the north-west. The RGB population in each of these field
is narrower than found in the disk annulus, and seems to lack the
population of redder RGB stars that are found in the disk field.

Figure~10 further quantifies the morphology of the RGB populations. The
upper panel shows photometric metallicity distribution functions
(MDFs) for the combined halo fields (H1 and H2; dot-dashed histogram),
disk annulus field (dashed histogram), and the combined substructure
fields (S1, S2 and S3; solid histogram).  The histograms are scaled to
have the same area. The distribution functions have been created by
comparing each star with $i_0 \le 23.5$ to 12Gyr, $[\alpha/Fe] = 0$,
isochrones, in the range $-2.5 \le [Fe/H] \le -0.2$\,dex, shifted to the
distance modulus of M33. A bilinear interpolation is conducted in
color-magnitude space to find the metallicity of the isochrone that
best matches with the star's position. These MDFs are uncorrected for
foreground contamination, which is clearly a significant contributor
at higher ($[Fe/H] \gtrsim -0.7$\,dex) metallicity. It is for this
reason that we do not include isochrones more metal-rich
than $[Fe/H] = -0.2$\,dex, since here contamination from MW disk dwarfs is
severe and, as Figure~10 demonstrates, few M33 RGB stars are this
metal-rich anyway. 

Error bars in the top panel of Figure~10 quantify the robustness of the
MDFs to photometric errors. The position of each star in the CMD is
modelled as a two dimensional Gaussian, centered at the colour and
magnitude of the star and with widths equal to the
photometric uncertainties. We calculate the MDF for each component a
total of 1000 times, where in each realisation we select the colour
and magnitude of each star from its distribution and conduct the
interpolation as standard. The error bars show the standard deviation
in each bin over the 1000 realisations. 

The lower panel of Figure~10 shows the foreground corrected MDFs for
the disk annulus field and the combined substructure field (dashed and
solid histograms, respectively). Here, the putative halo MDF has been
subtracted, scaled in order to match the number of bright stars ($i
\le 21$\,mags) in the disk/substructure field. Since bright stars are
mostly foreground stars, then this scaling accounts for possible
variation in the foreground level across the survey region. To first
order, any weak M33 halo signature in the disk and substructure fields
will also have been removed. Note that the larger fields, A and B,
could also be used for subtraction here, but these are unlikely to
contain many M33 halo stars given their larger distances from the galaxy.

With the usual caveats regarding the age and homogeneity of the
stellar populations, Figure~10 implies that the mean (median)
metallicity of the M33 substructure is $[Fe/H] \simeq -1.55$\,dex
($-1.6$\,dex), and is indicated in Figure~10 by the dotted
(dot-dashed) line. The interquartile range for the distribution is
$IQR = 0.5$\,dex, which will include a slight broadening due to
photometric uncertainties that is small in comparison to systematic
uncertainties. While the disk annulus has stars that are as metal-poor
as those found in the substructure, and has a similar IQR ($0.6$\,dex)
its mean (median) metallicity is notably higher, with $[Fe/H] \simeq
-1.25$\,dex ($-1.2$\,dex). Also, while the peak metallicity of the
substructure is close to its mean value, the peak metallicity of the
disk is higher still, at $[Fe/H] \sim -0.95$\,dex. Note that adopting
a 9\,Gyr age assumption would act to increase our metallicity
estimates by $\sim 0.2$\,dex.

Finally, Figure~11 shows the background-corrected Hess diagrams
corresponding to the MDFs shown in the lower panel of Figure~10. Here,
the Hess diagram corresponding to the combined ``halo'' fields has
been subtracted from the disk and combined substructure CMDs, using
the same scaling as for the MDFs. The pixels used in Figure~11 are
$0.05\times0.05$\,mags, and for clarity only positive residuals are
shown on a linear scale. A 12Gyr, $[\alpha/Fe] = 0.0$, [Fe/H] =
-1.2\,dex isochrones (the mean metallicity of the disk field) is
overlaid in both panels for reference. Figure~11 also shows that the
stars in both the disk and substructure regions extend to equivalently
blue (metal-poor?)  colours, and that the main difference between the
populations is at the red (metal-rich?) end.

\subsection{Surface brightness distribution of M33 substructure}

We now investigate in more detail the surface brightness distribution
of the M33 substructure. The previous MDF analysis suggests that,
while RGB stars of all photometric metallicities are present in the
structure, the majority have $-2.0 \le [Fe/H] \le -1.0$\,dex
(Figure~10). We therefore adopt this cut to enhance the substructure
relative to other features in our survey, and we continue to use the $i_0
\le 23.5$ magnitude cut.

\subsubsection{Background variations}

Figure~12 investigates how the Galactic foreground varies as a
function of Galactic latitude for stars that coincide with our
``optimal'' RGB metallicity cut of $-2.0 \le [Fe/H] \le -1.0$\,dex and
$i_0 \le 23.5$. Here, we have plotted the density variation with
Galactic latitude of stars that satisfy this cut and which are located
in the 1 degree wide strips highlighted in Figure~8 as ``A'' and
``B'', centered at $l = 135.7^\circ$ and $l = 130.9^\circ$,
respectively. These strips were purposely positioned away from the
substructure on either side of our survey to probe large-scale
variations in the foreground. Circular points in Figure~12 correspond
to the density distribution in strip A and triangles correspond to
strip B.

Figure~12 shows that the variation in the foreground over our survey
area in our optimal RGB metallicity cut is relatively stable. In
particular, there are no significant trends with Galactic latitude -
the density of counts do not vary systematically between $b =
-29^\circ$ and $b = -33^\circ$. In addition, the counts in strip A and
strip B, separated by nearly $5^\circ$ in Galactic longitude, are
approximately the same; dotted lines in Figure~12 indicate the mean
and $1\sigma$ variations in density for strip A, and dashed lines
indicate the same for strip B. There is little difference between the
means, and they lie within the $1\sigma$ variation. As such, we
conclude that a constant background is a suitable approximation for
the subsequent surface brightness analysis.

\subsubsection{Stellar density - surface brightness}

Figure~13 is the stellar density map for our chosen optimal RGB
metallicity cut ($-2.0 \le [Fe/H] \le -1.0$\,dex and $i_o \le 23.5$),
constructed in an identical way to Figure~8, and shown with linear
scaling. The grey contour level shows the $1\sigma$ detection
threshold in star counts and the black contours show $2, 5, 8$ and
$12\sigma$ detection thresholds. The substructure is seen clearly in
this image, and extends out nearly as far as the $50$\,kpc boundary of
the survey (shown with a dashed line). These low level contours also
suggest that the southern extension of the substructure may extend out
as far as its northern counterpart, although at large radius it is a
lower surface brightness, less coherent, feature than in the north.

Figure~14 shows background-corrected radial profiles for M33,
constructed in 1 degree wide strips, in two orthogonal directions,
using identical cuts as for Figure~13. The top panel shows the radial
profile in an ``on-stream'' region, and the bottom panel shows the
profile in an ``off-stream'' region. Each bin is chosen so that the
signal-to-noise due to Poisson errors in that bin (prior to background
subtraction) is at least 10, and the displayed error bars additionally
include the uncertainty in the background. Here,
$\theta$ refers to the angle, measured east-from-north, from the
northern semi-major axis of M33. Dashed lines correspond to the radius
of the $\mu_B = 25\,$mag\,arcsec$^{-2}$ isophote. The substructure is
clearly visible in the top panel of Figure~14 as an extended,
slowly-declining, component. A north-south asymmetry is present, with
the northern component more dominant out to larger radius than in the
south.  This feature is also present in the ``off-stream'' region,
particularly in the east, but does not extend as far as in the
``on-stream'' regions. Beyond $1 - 1.5$degrees in the
``off-stream'' region, there are hints of a slowly-declining
component. Presumably this can be attributed to an extended halo in
M33, also hinted at in the ``halo'' CMDs in Figure~9. However, other
interpretations may exist, and we will return to this topic in a later
paper.

Figures~13 and 14 show density distributions of a subset of candidate
RGB stars; to convert this into surface brightness necessarily
requires some assumption about the luminosity function to correct for
the unresolved stellar component. In Section 3.3, we showed that the
resolved populations in the substructure are consistent with a
relatively metal-poor ($[Fe/H] \simeq -1.5$\,dex) population, similar to
that of some of the dwarf spheroidal (dSph) galaxies around M31 (e.g.,
\citealt{dacosta1996,dacosta2002,mcconnachie2005a}). The two brightest
dSph galaxies that completely lie within the first-year PAndAS
footprint are Andromeda I and III, and these have independent
measurements of their total luminosity based on unresolved light
(\citealt{mcconnachie2006a}). As such, these dSphs can provide us with
an observationally-motivated transformation between RGB star counts
and integrated luminosity. Implicit in this technique is that the
luminosity function of the substructure is the same as for Andromeda I
(III).

Our adopted distance modulus for M33 of $\left( m - M \right)_o =
24.54 \pm 0.06$ corresponds to a magnitude at the tip of the RGB of
$I_o = 20.49$ (\citealt{mcconnachie2005a}). In the CFHT filter system,
this corresponds to $i_o = 20.97$ (see the transformations given in
\citealt{ibata2007}) for a mean color at the tip of the RGB of
$\left(V - I\right) \simeq 1.7$ (e.g.,
\citealt{mcconnachie2004a}). Thus Figure~13 shows stars within the top
2.5 magnitudes of the RGB population in M33, given our cut-off
magnitude of $i_o = 23.5$.

\cite{mcconnachie2006a} measure integrated luminosities of $M_V^I =
-11.8 \pm 0.1$ and $M_V^{III} = -10.2 \pm 0.3$ for Andromeda~I and
III, respectively. Using these galaxies' half-light radii, ellipticity
and position angles, as measured by the same authors, we record the
total luminosity of stars observed in the PAndAS data within 2.5\,mags
of the tip of the RGB and within one half-light radius from each of
these galaxies. We correct this value for foreground by comparison to
nearby regions, and we multiply the final value by two to obtain the
total ``RGB luminosity'' of Andromeda I and III. We require a 2.30 and
2.18 magnitude offset for Andromeda I and III, respectively, to
correct the RGB luminosity to the total integrated luminosity. This
implies that the top 2.5 magnitudes of the RGB are contributing $\sim
10 - 15\,\%$ of the total light of these galaxies, which seems
reasonable for an old population (e.g., Figure~2 in
\citealt{renzini1988}).

To convert Figure~13 to a background-corrected surface brightness
scale, we measure the total luminosity of all the stars in each pixel
of the unsmoothed version of Figure~13, and subtract the
background. We correct this luminosity for RGB stars that are not
included in our metallicity cut (Figure~10 shows that 69\% of RGB stars
with $i_o \le 23.5$ are included in our optimal metallicity cut, $-2.0
\le [Fe/H] \le -1.0$\,dex). We then smooth the map as before, and
correct the value of each pixel by 2.30 (2.18)\,magnitudes using the
conversion for Andromeda I (III).

Figure~15 plots the relation between the number of RGB stars in each
pixel of Figure~13 and the integrated luminosity of the pixel as
derived above (using the Andromeda I transformation). Arrows on the
$x$-axis show the 1, 2, 5, 8 and $12\sigma$ number density contour
levels from Figure~13. We fit a simple linear relationship to the data
points in Figure~15 by binning the data above $n_{RGB} = 500$\,stars
degree$^{-2}$ and use this relationship to calculate the surface
brightness values corresponding to our contour levels. These are listed in
full in Table~1, for both the Andromeda I and III transformations.
The right vertical axis of Figure~15 shows the surface brightness
scale for the Andromeda~I transformation. Note that, at low stellar
density, this relationship overestimates the actual ``observed''
luminosity density, since star counts and luminosity are only in
direct proportion if the luminosity function is well sampled. At low
star count levels, more luminous stars, that are intrinsically more
rare than less luminous stars, are undersampled. Thus regions with low
star counts are found to be underluminous compared to expectations
from regions with higher star counts. To illustrate this, we fit a
second linear relationship to Figure~15 using only regions with
$n_{RGB} \le 350$\,stars degree$^{-2}$. The surface brightness scale
that this produces is listed in Table 1 in parentheses.

The background-subtracted surface brightness levels listed in Table~1
clearly have large systematic uncertainties. These include the unknown
form of the true luminosity function; the uncertainty in the
integrated luminosity of the system we are normalizing to (Andromeda~I
and III); the stochastic variations in stellar density/stellar
luminosity in sparse regions. The spread in surface brightness values
in each row of Table 1, combined with the uncertainties in total
luminosities of Andromeda~I and III, suggests that the zero-point of
the surface brightness scale is accurate to the level of $\pm
0.5$\,mags\,arcsec$^{-2}$.

\subsubsection{Integrated light}

We integrate the surface brightness map to calculate the total
luminosity of the substructure surrounding M33. Defining the edges of
the substructure by the $2\sigma$ and $12\sigma$ contours in Figure~13
results in a total luminosity of $M_V = -12.7 \pm 0.5$ after
correction to the distance modulus of M33. This number is a lower
limit since it assumes there is a hole in the middle of the
substructure delineated by the $12\sigma$ contour. If instead we allow
for an additional constant contribution to the flux from within this
area equal to 30.8\,mag\,arcsec$^{-2}$ (see Table~1), then the
luminosity of the substructure increases by $\sim 0.4$\,mags. Thus the
total luminosity of this feature is of order $\sim 1\%$ of the total
luminosity of M33. For comparison, the Fornax dwarf galaxy - one of
the brightest dwarf spheroidal galaxies in the Local Group - has a
luminosity of $M_V = -13.2$ (\citealt{mateo1998a}).

\subsection{Distribution of young stars} 

The top panel of Figure~16 shows a tangent plane projection of the
position of all blue stars with $(g - i)_o < 0$ and $i_o <
23.5$\,mags, that predominantly selects bright main sequence and
blue-loop stars, indicative of a young population. Foreground
contamination in this color interval is minimal. An annulus sampling
stars in the outer disk is shown, where the inner edge corresponds to
the radius at which $\mu_B = 25$\,mags\,arcsec$^{-2}$, and the outer
edge is 1.25 times further out. Circular points in the lower panel of
Figure~16 show the variation in stellar density around this annulus as
a function of $\theta$, defined earlier. Error bars show the Poisson
uncertainties. Two clear overdensities are visible, at $\theta \simeq
130^\circ$ and $\theta \simeq310^\circ$. It is notable that these
overdensities of young stars are separated by $180^\circ$, and that
they have a similar orientation to the RGB overdensity and the HI
warp. Thus it is not just the gas and {\it evolved} stars in M33 that
have a distorted distribution, and it is perhaps not surprising that
the young stars appear to follow the same distorted distribution as
the underlying gas.

\section{Discussion}

\subsection{Revisiting detections of the M33 stellar halo}

\cite{ibata2007} measure the radial surface brightness of M31 in the
direction of M33. The M31 surface brightness is shown to decrease to
$\sim 33$\,mags\,arcsec$^{-2}$ at $\sim 11^\circ$ ($\sim 165$\,kpc)
from M31 before it starts increasing again approaching M33. This is
attributed to a detection of M33's stellar halo. However, with the
completion of azimuthal coverage around M33 presented herein, it is
clear that \cite{ibata2007} had in fact detected the north-eastern
extension of the stellar structure examined here, and not an extended,
bona-fide, halo population.

\cite{mcconnachie2006c} analyze Keck/DEIMOS spectroscopy of giant
stars in two fields on the southern minor axis of M33, 38\,arcmins
(9\,kpc) from the center of the galaxy. The dominant component visible
in the kinematics for this field is the disk. In addition, they also
identify a population of stars at more negative velocities than the
systemic velocity of M33 ($v_h = -179$\,km\,s$^{-1}$). These stars are
consistent with belonging to a pressure supported stellar halo, with a
dispersion of $\sigma \sim 50$\,km\,s$^{-1}$.  However, in light of
the stellar substructure identified in the PAndAS data, the analysis
by \cite{mcconnachie2006c} needs to be revisited. Their interpretation
of the M33 stellar kinematics was made with reference to a simple
model of a regular, unperturbed, stellar disk with a rotation curve
similar to that observed in the HI (\citealt{corbelli2003}). Using a
more complex model that accounts for the observed substructure could
help provide an explanation for the kinematic structure that
\cite{mcconnachie2006c} observe.

Other authors have claimed M33 halo detections; \cite{chandar2002} and
\cite{sarajedini2006} present evidence for distinct populations of
star clusters and RR~Lyrae, respectively, in M33. For the former, the
older clusters are suggested to belong to two kinematic components (a
disk and a halo), whereas the younger clusters are consistent with
belonging to only one component (a disk). For the latter, examination
of the reddening and period distribution of individual stars suggests
a bimodal population, with the less extincted, longer period (more
metal-poor) stars attributed to a halo population. These studies are
based on observations at much smaller radii than probed here ($<
0.5^\circ$ from the center of M33), and their interpretations are
unlikely to be affected by the presence of this new substructure.

\subsection{Origin of the M33 substructure}

\subsubsection{An accreted dwarf galaxy?}

The luminosity of the structure surrounding the M33 disk is at least
$M_V = -12.7 \pm 0.5$, similar to the integrated luminosity of a
bright Local Group dwarf spheroidal, such as Fornax or
Andromeda~II. Further, with an estimated median photometric
metallicity of $[Fe/H] = -1.6$\,dex, it would be consistent with the
broad luminosity - metallicity correlation that is observed for dwarf
galaxies. If the M33 substructure is a tidal stream from a dwarf
galaxy, then it is reasonable to ask if the progenitor is visible.

Only one dwarf satellite has been observed in the PAndAS footprint
around M33 (excluding the previously known Andromeda~II dwarf galaxy
at the edge of the survey region), the low luminosity Andromeda~XXII
($M_V \simeq -6$; \citealt{martin2009}). There is no reason to suspect
it as the progenitor of the substructure, given that it lies far out
near the southern minor axis away from the main body of the
feature. Likewise, Andromeda~II shows no connection to the
substructure, and anyway it is a relatively bright galaxy that does not
show any sign of distortions (e.g., \citealt{mcconnachie2007a}). We
cannot rule out the possibility that either the putative progenitor is
entirely destroyed, or that it lies directly in front or behind the
main disk of M33.

\subsubsection{A tidal distortion due to M31?}

Our favored interpretation of the large substructure in M33 is that
it is due to a a tidal interaction with M31. A preliminary exploration
of this idea was presented in \cite{mcconnachie2009b}, and Dubinski et
al. (2010, in preparation) develop this model. Here, we summarize a
few salient points.

\subsubsubsection{Morphology}

Perhaps the most convincing piece of evidence for a tidal origin is
the morphology of the feature. Although slightly asymmetric, it has a
classical ``S-shaped'' appearance typical of a system that is tidally
interacting with a more massive companion, and it resembles an
extremely warped stellar disk. While the disruption of a dwarf galaxy
is plausible, most orbits for the putative progenitor would not
naturally produce a structure with the observed morphology.

\cite{rogstad1976} first demonstrated that the HI disk of M33 was
highly warped, and this has since been verified by \cite{corbelli1997}
and \cite{putman2009}. As such, it is perhaps
unsurprising to find that the stellar disk is similarly
perturbed. M. Putman kindly provided us with her group's integrated HI
map of M33 observed as part of the Galactic Arecibo L-Band Feed Array
HI (GALFA-HI) Survey. Following Figure~1 of \cite{putman2009}, the
left panel of Figure~17 shows the integrated HI contours of M33, where
contour levels are $8.3 \times 1.5^n \times 10^{18}$\,cm$^{-2}$
($n=0...13$). The middle panels shows a simple model of the disk of
M33, using the variation in position angle and inclination derived by
\cite{corbelli1997}. Here, we have modeled the density of the disk in
three dimensions by an exponential-sech$^2$ function, and integrated
along the line of sight. Contours represent one magnitude drops in
surface brightness from the maximum surface brightness of the
disk. The right panel shows the observed surface brightness
distribution map of M33, previously shown in Figure~13. Figure~17
shows that the stars and gas warp in the same direction, and both are
reasonably well matched by the simple model of
\cite{corbelli1997}. While such a similarity could be coincidence, it
is also qualitative evidence for a common origin.

\subsubsubsection{Metallicity}

The analysis in Figure~10 shows that the MDF of the disk stars in the
D1 annulus has a broadly similar IQR as the stellar
substructure. However, the latter has a higher mean metallicity - by
about 0.3\,dex - than the disk population, and the peak metallicities
of the two populations differ by $\sim 0.6$\,dex. Comparison of the
two Hess diagrams in Figure~11 shows that the two RGB populations are
similar at the blue end but that the disk population has a significant
redward extension that is absent in the substructure population. Given
these differences, can the substructure be related to the disk
population?

It is difficult to reconcile the differences between the CMDs and
photometric MDFs of the substructure and disk fields with the data
presented herein. We note that metallicity gradients have been
observed in the M33 disk using a range of tracers (e.g., AGB stars,
\citealt{cioni2009}; PNe, \citealt{magrini2009}; Figure~20 of
\citealt{barker2007} and references therein). \cite{tiede2004} measure
a gradient of $\Delta$[Fe/H]$/\Delta R_{proj} = -0.06 \pm
0.01$\,dex\,kpc$^{-1}$, based on the color of RGB stars, but so far no
studies have probed the disk at radii close to that of the
substructure studied here ($\sim 12$ disk scale-lengths, using the
estimate by \citealt{regan1994}, measured in the $J$ band). This is
the very outer, very low surface brightness part of the disk, where
few if any constraints exist on the intrinsic population of disks. If
an interaction has taken place, then this may well act to radially
redistribute stars and gas, which will introduce and/or reshape
gradients in age and metallicity, further complicating
matters. Systematic gradients in age are particularly difficult for
photometric-based analyses to account for, and spectroscopic
metallicity studies will be required to better understand the
population differences that this photometric study reveals.

\subsubsubsection{Other considerations}

The preliminary models of a tidal encounter between M31 and M33
presented in \cite{mcconnachie2009b} demonstrate that a relatively
close interaction between M31 and M33 can plausibly produce a stellar
feature on the scale observed, while also satisfying the known
phase-space constraints of these galaxies (including the proper motion
of M33; \citealt{brunthaler2005}). Such an interaction may also affect
other tracers in the outer regions of M33. For example, M33 appears to
lack a significant population of globular clusters at larger radii;
\cite{huxor2009} discover four M33 globular clusters beyond the $\mu_B
= 25$\,mag\,arcsec$^{-2}$ isophote, bringing the total beyond this
radius to only five (including the extended remote globular cluster
discovered by \citealt{stonkute2008}). All but the least remote of
these clusters are in the hemisphere of M33 furthest from M31, which
\cite{huxor2009} suggest may be tentative signs of tidal disturbance
in M33 at large radius.

Another useful tracer of galaxies at large radii are planetary nebulae
(PNe). \cite{ciardullo2004} conduct a survey of the PNe population of
M33 over the entire main disk of M33. They find that the velocity
dispersion does not decrease over $\sim 4$ disk scale-lengths and has
a value of nearly $\sigma \simeq 20$\,km\,s$^{-1}$ at $\sim10$\,kpc
from the center of the galaxy. One possible explanation for this
behavior is that the outer PNe have been dynamically heated due to
tidal interactions with M31. Similarly, the study by M06 measure a
velocity dispersion of the M33 disk at $\sim 9$\,kpc of $\sim
12.5\,$km\,s$^{-1}$, higher than expected from simple
dynamical arguments, and again consistent with the idea of tidal
heating. Finally, \cite{putman2009} also measure high velocity
dispersion for the HI across the disk of M33, which they suggest may
be consistent with a recent interaction with M31.

\section{Summary}

In this paper we have presented an analysis of $g$ and $i$ photometry
of approximately 40 square degrees surrounding M33, taken as part of
the {PAndAS program using the MegaPrime/MegaCam instrument on CFHT. We
  show that the large stellar structure surrounding M33, first
  identified by \cite{mcconnachie2009b}, is broadly aligned with the
  warped HI disk of M33
  (\citealt{rogstad1976,corbelli1997,putman2009}).  We trace this
  feature to $\mu_V \simeq 33$\,mags\,arcsec$^{-2}$ and show that it
  extends to projected radii of $\sim 40$\,kpc from the center of M33
  in both the north-west and south-east quadrants of the galaxy,
  giving the structure an ``S'-shaped'' appearance. Its stellar
  populations are consistent with an old population with $<[Fe/H]>
  \sim -1.6$\,dex and an interquartile range in metallicity of
  $0.5$\,dex. We calculate a lower limit to the integrated luminosity
  of the structure of $-12.7 \pm 0.5$\,mags, comparable to a bright
  dwarf galaxy such as Fornax or Andromeda~II and slightly less than
  $1\%$ of the total luminosity of M33. We show that there is
  tentative evidence for a distortion in the distribution of young
  stars near the edge of the HI disk that occurs at similar azimuth to
  the warp in HI. These data also hint at an extended halo component
  that dominates at very large radius and which will be explored in
  detail in a forthcoming contribution. We revisit studies of M33 and
  its stellar populations in light of these new results, and we
  discuss possible formation models for the vast stellar structure.
  Our favored scenario is the tidal disruption of M33 as it orbits
  around M31.

  \acknowledgements Based on observations obtained with
  MegaPrime/MegaCam, a joint project of CFHT and CEA/DAPNIA, at the
  Canada-France-Hawaii Telescope (CFHT) which is operated by the
  National Research Council (NRC) of Canada, the Institute National
  des Sciences de l'Univers of the Centre National de la Recherche
  Scientifique of France, and the University of Hawaii. We would like
  to thank the entire staff at CFHT for their great efforts and
  continuing support throughout this project. We thank Mary Putman for
  supplying us with the integrated HI column density map for M33 and
  for comments on the manuscript. Thanks to all our collaborators in
  the PAndAS, particularly Scott Chapman, Mark Fardal, Nicolas Martin
  and Jorge Pe{\~n}arrubia, for a careful reading of the manuscript
  and thoughtful comments, and finally to the anonymous referee, for
  useful comments that improved this paper. A.M.N.F. acknowledge
  support by a Marie Curie Excellence Grant from the European Commission under con- tract MCEXT-CT-2005-025869.

\clearpage
\newpage

\begin{figure*}
  \begin{center}
    \includegraphics[angle=270, width=12cm]{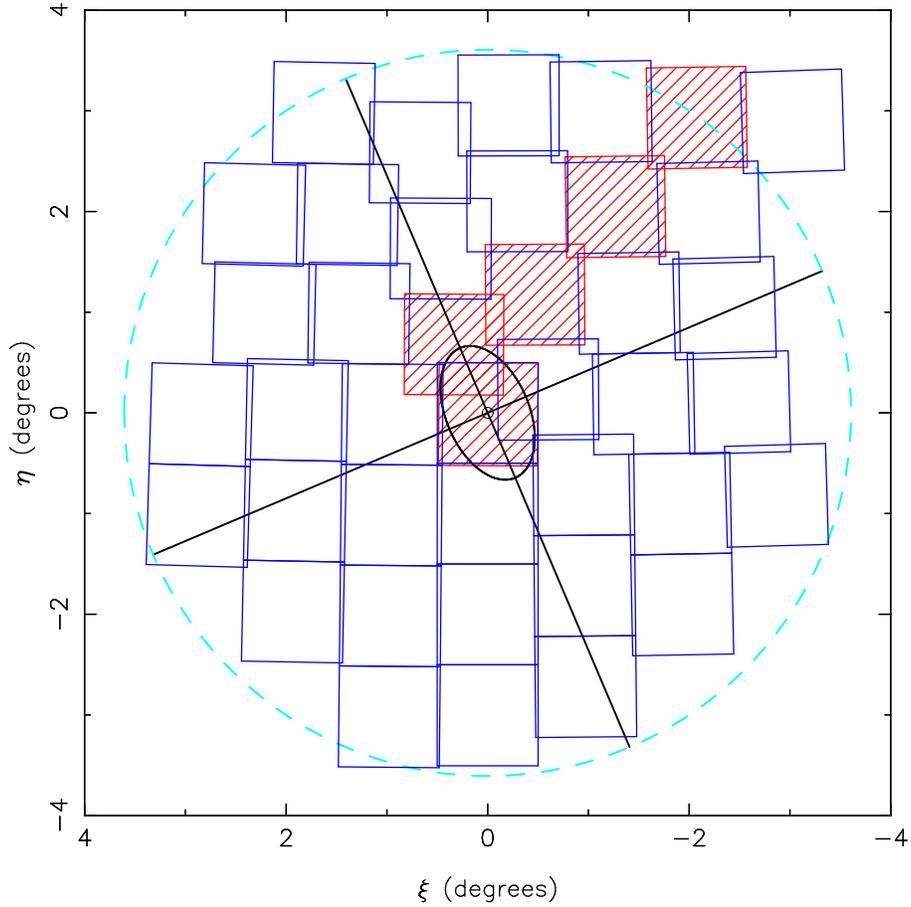}
    \caption{A tangent plane projection of the location of 40 PAndAS
      fields in the vicinity of M33. The ellipse marks the radius at
      which $\mu_B \simeq 25$\,mag\,arcsec$^{-2}$. Major and minor
      axes are shown. The dashed circle corresponds to a projected
      radius of 50\,kpc from M33. Hatched fields were presented in
      \cite{ibata2007}. Open fields were observed in S08B. }
  \end{center}
\end{figure*}

\clearpage
\newpage

\begin{figure*}
  \begin{center}
    \includegraphics[angle=270, width=12cm]{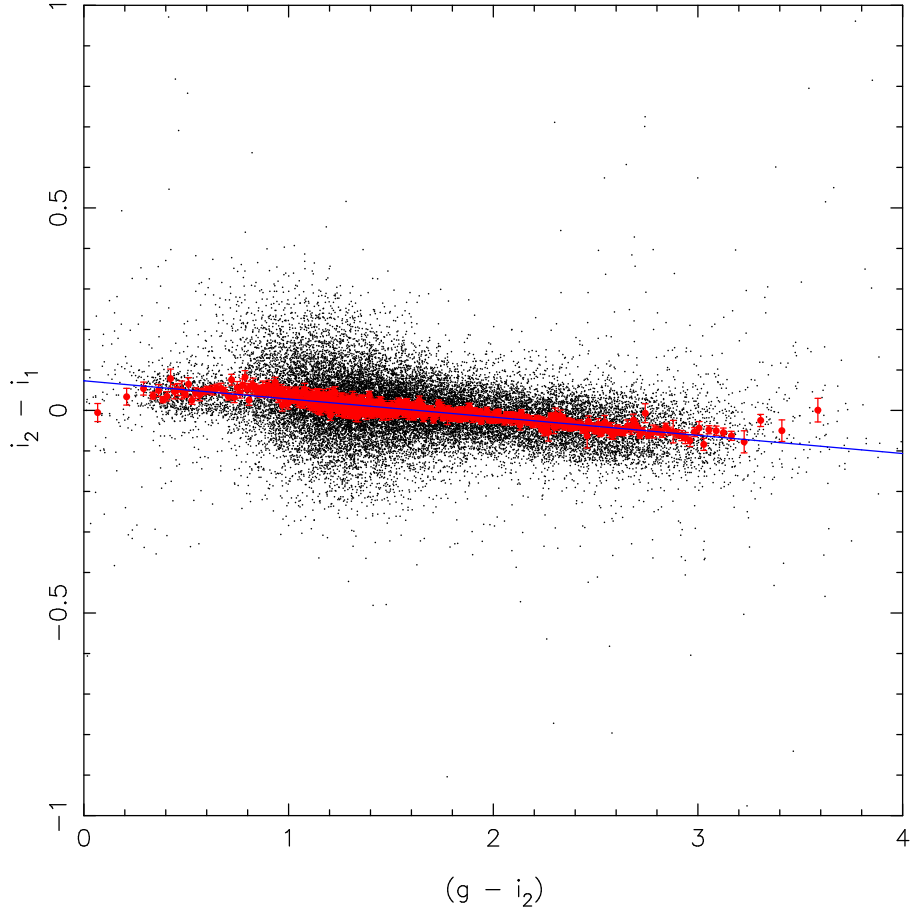}
    \caption{The difference between the old ($i_2$) and new ($i_1$)
    CFHT/MegaPrime $i$ filter (pre- and post-2007, respectively) as a
    function of color $(g - i_2)$ using stellar sources in regions
    where we have overlap between the two generations of
    $i-$filters. Red points and their error bars show the mean and
    variance in bins containing 50 stellar sources. The straight line
    shows the best fit linear relation, described by Equation~1.}
  \end{center}
\end{figure*}

\clearpage
\newpage

\begin{figure*}
  \begin{center}
    \includegraphics[angle=270, width=12cm]{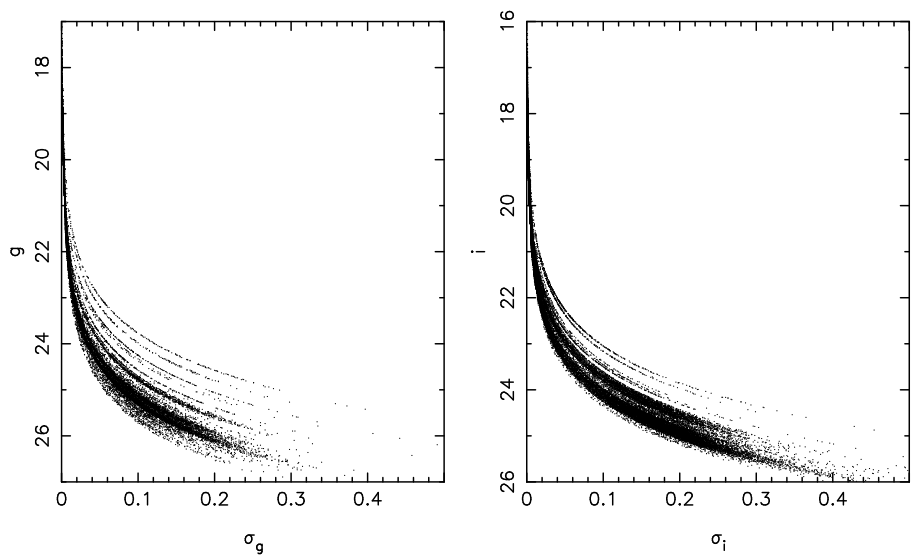}
    \caption{Photometric errors as a function of magnitude. The
    distinct sequences are due to fields that were
    taken in slightly different observing conditions. Generally, our
    photometry is accurate to better than 0.2\,mags for $i \le 24.5$
    and $g \le 25.5$.}
  \end{center}
\end{figure*}

\clearpage
\newpage

\begin{figure*}
  \begin{center}
    \includegraphics[angle=270, width=12cm]{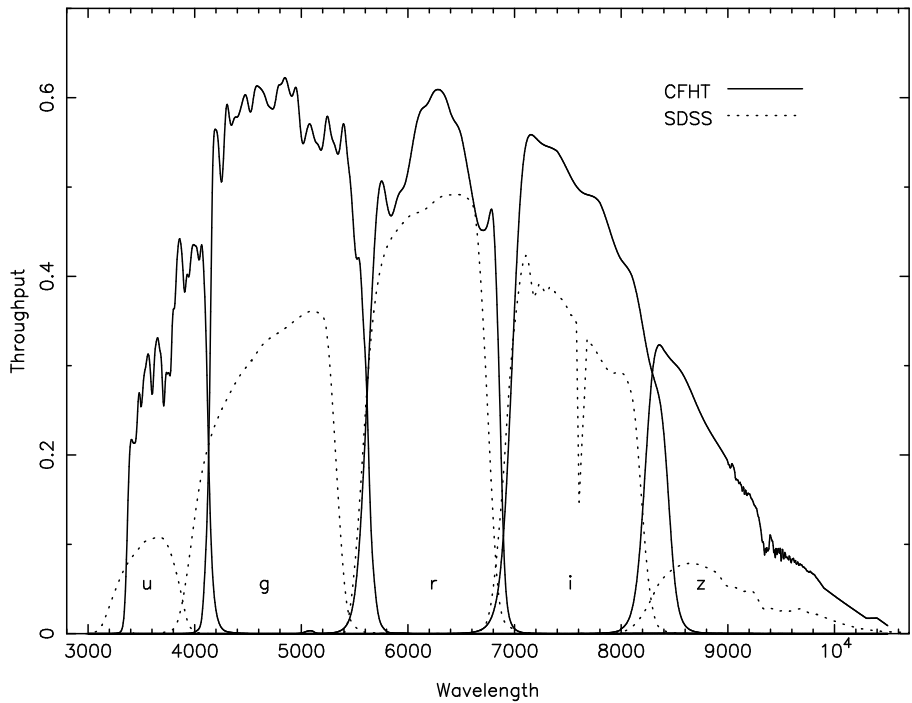}
    \caption{A comparison of the SDSS and CFHT filter transmission
      curves, taking into account the throughput of the telescope
      optics, the quantum efficiency of the CCDs, and the atmospheric
      transmission curves. The CFHT $g$ filter is shifted to the red in
      comparison the SDSS equivalent, and the CFHT $i$ filter is
      broader than the SDSS equivalent by $\sim 250\AA$ towards the
      red end of the spectrum.}
  \end{center}
\end{figure*}

\clearpage
\newpage

\begin{figure*}
  \begin{center}
    \includegraphics[angle=270, width=12cm]{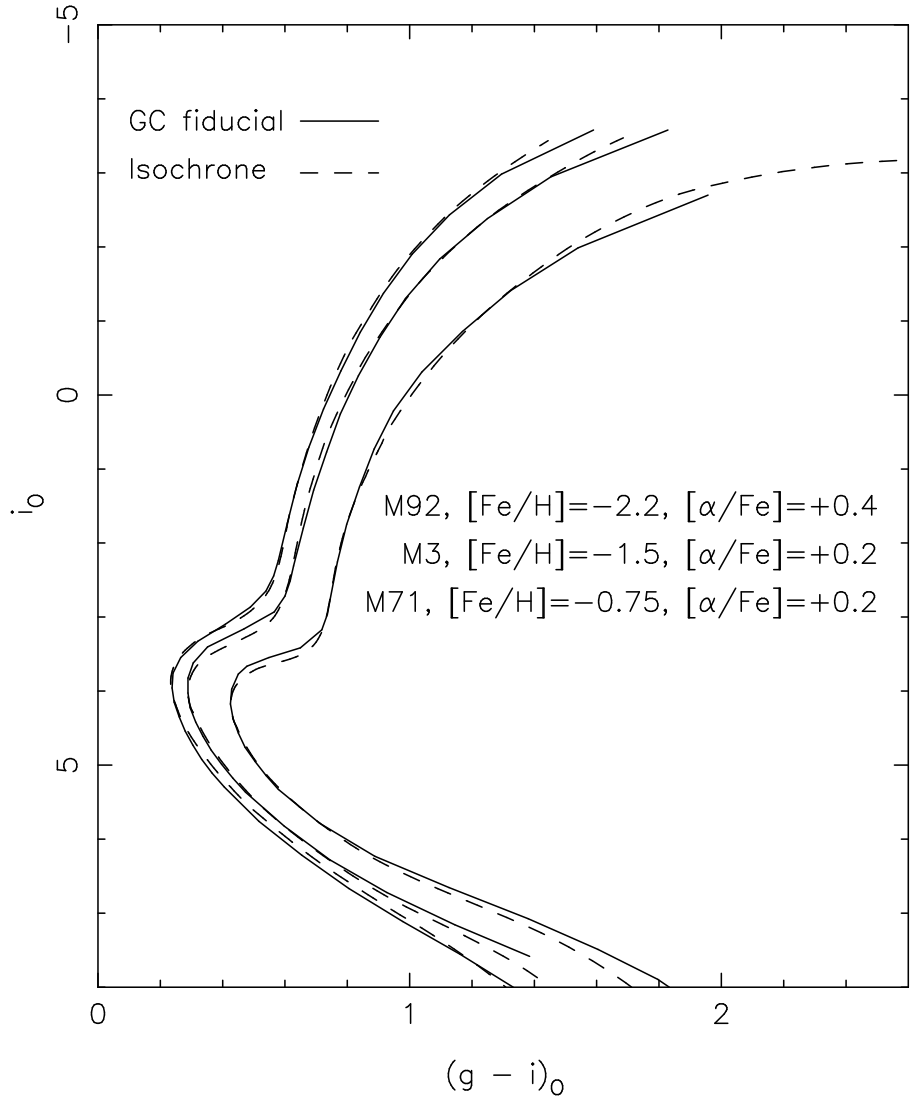}
    \caption{A comparison of the CFHT isochrones for 14Gyr populations
      (dashed lines) to fiducial sequences for the globular clusters
      M92, M3 and M71 taken from \cite{clem2008} (solid lines). The
      key denotes the [Fe/H] and [$\alpha$/Fe] used for each
      cluster. The assumed distance moduli and E(B-V) values for the
      clusters are given in the text. Clearly, the isochrones match
      the fiducial sequences very well. Only the giant branch of the
      isochrones will be required for the subsequent analysis.}
  \end{center}
\end{figure*}

\clearpage
\newpage

\begin{figure*}
  \begin{center}
    \includegraphics[angle=270, width=12.cm]{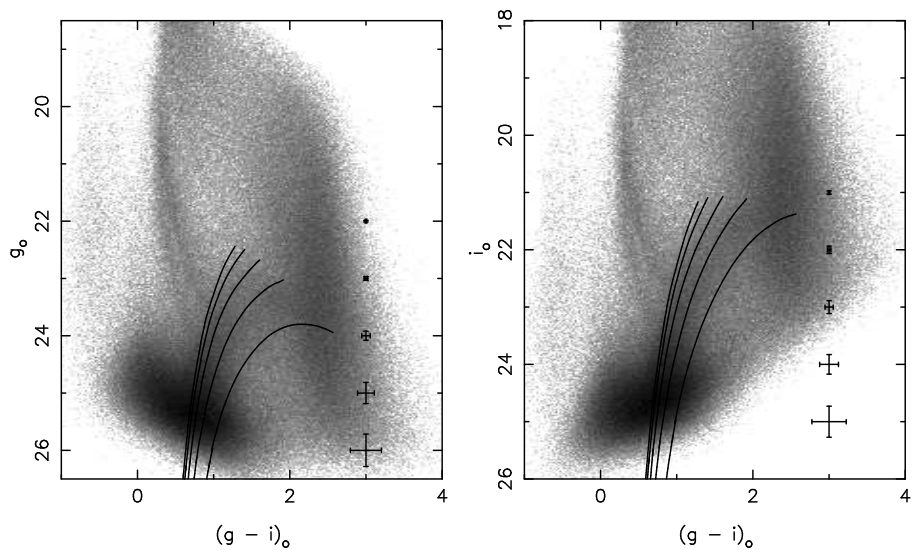}
    \caption{$g_o$ (left) and $i_o$ (right) color-magnitude (Hess)
      diagrams of all stellar objects in the M33 survey region shown
      in Figure~1, excluding the central field. Bins are $0.025 \times
      0.025$\,mags, displayed with a logarithmic scaling. Mean errors
      as a function of magnitude are shown. Overlaid are isochrones
      corresponding to a 12\,Gyr, [$\alpha/$Fe] $=0.0$, stellar
      population at the distance of M33 with $[Fe/H] = -2.5, -2, -1.5,
      -1$ and $-0.5$\,dex.}
  \end{center}
\end{figure*}

\clearpage
\newpage

\begin{figure*}
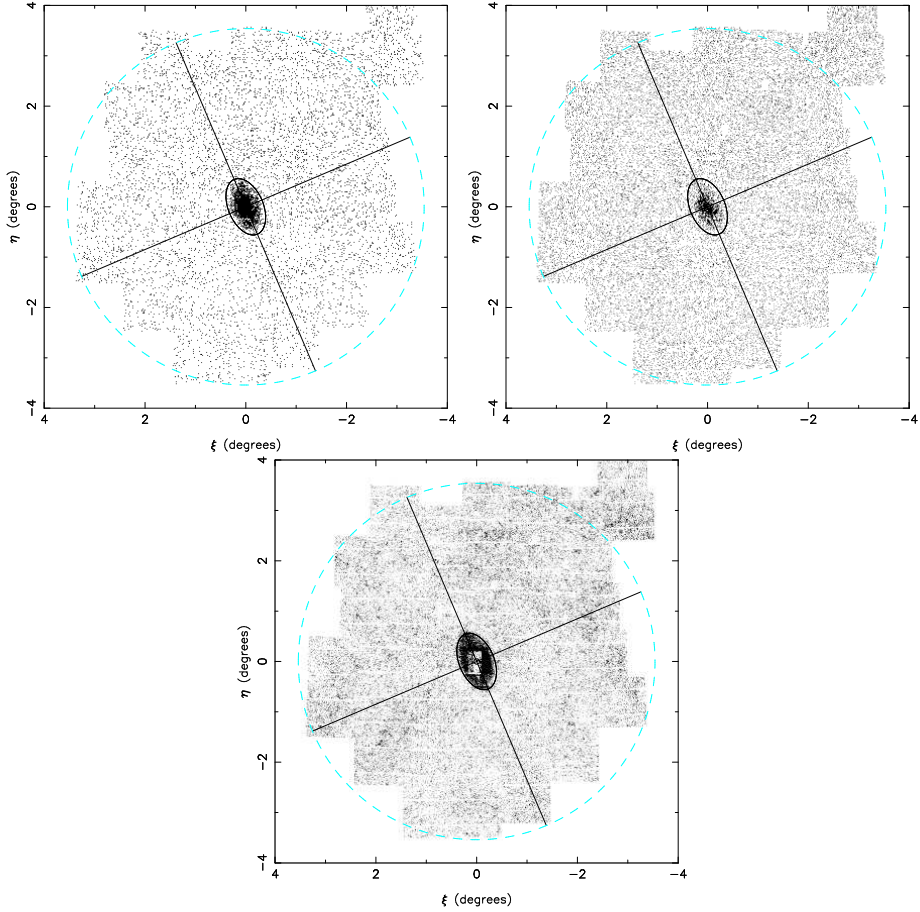

  \begin{center}
    \includegraphics[angle=270, width=6cm]{figure7a.ps}
    \includegraphics[angle=270, width=6cm]{figure7b.ps}
    \includegraphics[angle=270, width=6cm]{figure7c.ps}
    \caption{The spatial distribution of various contaminants in our
    photometric survey of M33. Maps use $1.2 \times 1.2$\,arcmins
    pixels and are displayed with a linear scale. Left panel: Galactic
    halo stars, selected to have $19.0 < i_0 < 22.0$, $0.1 < (g - i)_0 <
    0.6$. Middle panel: Galactic disk stars, selected to have $17.0 <
    i_0 < 20.0$, $1.5 < (g - i)_0 < 3.0$. Right panel: extended sources
    (galaxies) with $17.0 < i_0 < 23.5$. Faint, compact galaxies
    misidentified as stars are likely to have a similar spatial
    distribution to the latter population.}
  \end{center}
\end{figure*}

\clearpage
\newpage

\begin{figure*}
  \begin{center}
    \includegraphics[angle=270, width=12cm]{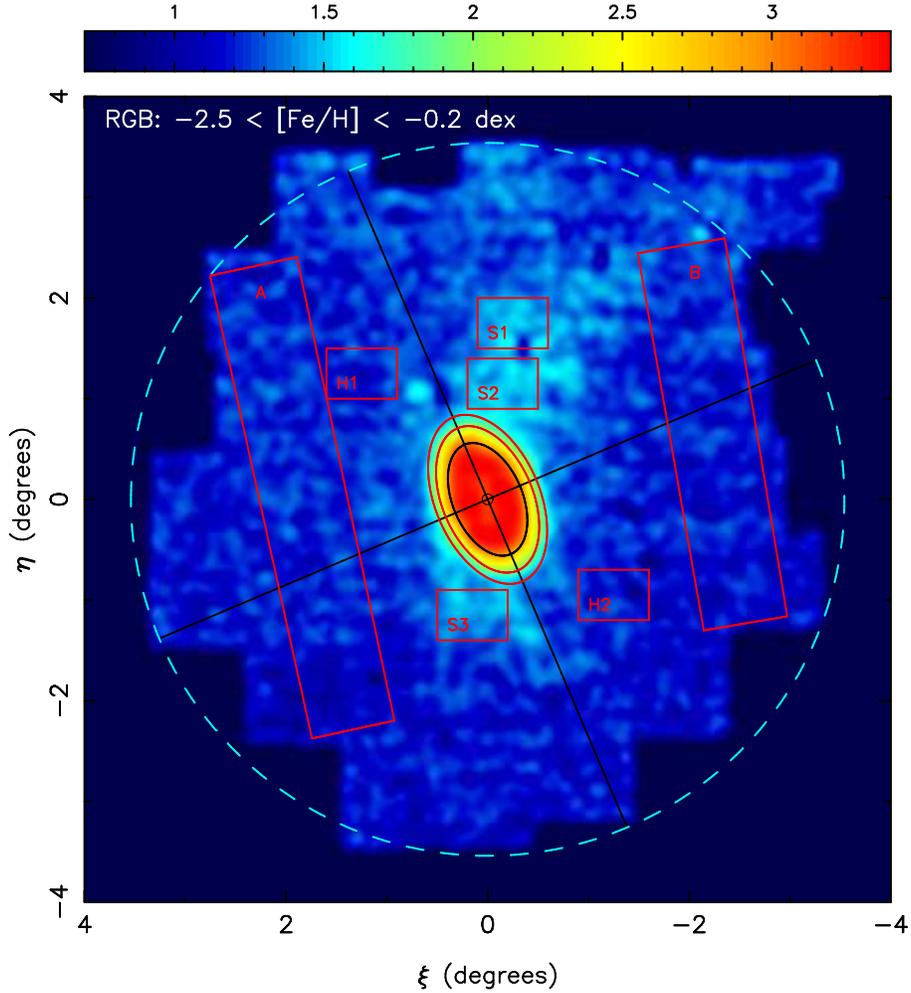}
    \caption{The spatial distribution of candidate RGB stars at the
    distance of M33 with $i_0 \le 23.5$ in a tangent-plane projection
    displayed with logarithmic scaling. These stars lie in the same
    color - magnitude locus as 12 Gyrs, $[\alpha/Fe] = 0.0$,
    isochrones, with metallicities between $-2.5 \le [Fe/H] \le
    -0.2$\,dex, shifted to the distance modulus of M33.  A slight
    gradient from north to south is visible, since the more metal-rich
    RGB stars overlap in color-magnitude space with the foreground
    disk population. The small red boxes and the red elliptical annulus
    each have an area of 0.35 square degrees and probe areas of
    interest around M33. The large rectangular strips, A and B, each
    probe 1 degree strips in Galactic latitude away from the main body
    of M33 and are used later to examine variations in the Galactic
    foreground (see text for details).}
  \end{center}
\end{figure*}

\clearpage
\newpage

\begin{figure*}
  \begin{center}
    \includegraphics[angle=270, width=15.cm]{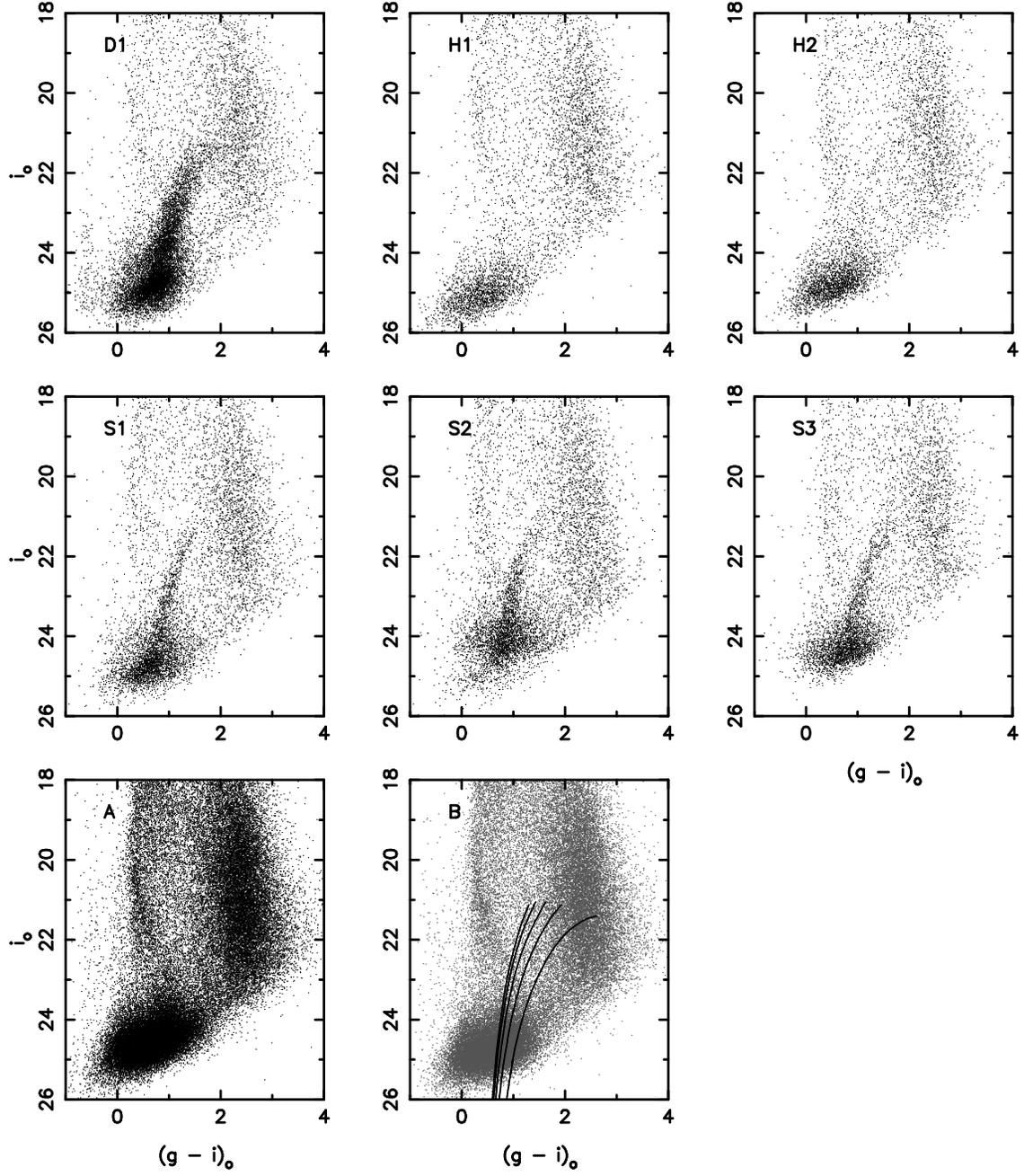}
    \caption{$i-$band CMDs of the areas highlighted in Figure~8. Field
      D1 (corresponding to the elliptical annulus) samples stars in
      the outer disk of M33; H1 and H2 sample areas away from obvious
      substructure and are putative ``halo'' fields; S1, S2 and S3
      sample the substructure around the disk of M33 at similar radii
      to the halo fields. Each of these field has the same area (0.35
      square degrees). CMDs for area A and B are also shown, and
      isochrones corresponding to [Fe/H] $= -0.5, -1.0, -1.5, -2.0,
      -2.5$\,dex are overlaid on area B for reference.}
  \end{center}
\end{figure*}

\clearpage
\newpage

\begin{figure*}
  \begin{center}
    \includegraphics[angle=270, width=12cm]{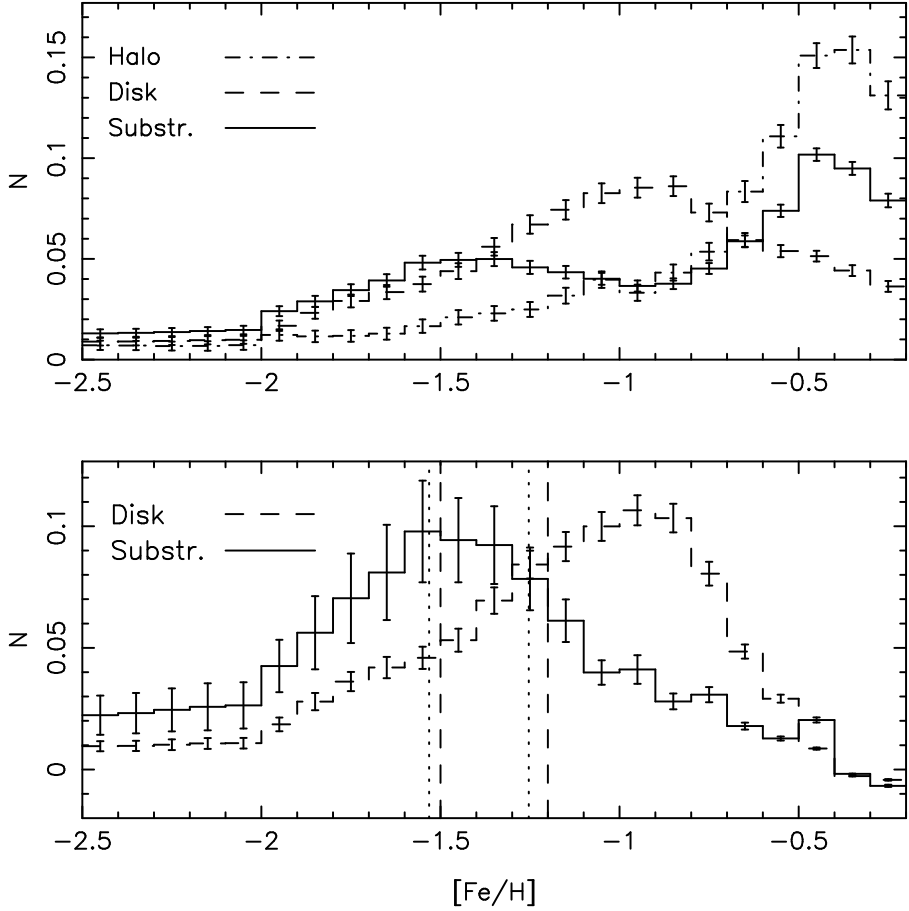}
    \caption{Upper panel: The photometric metallicity distribution
      functions (MDFs) for the combined putative halo fields (H1 and
      H2; dot-dashed histogram), disk annulus field (D1; dashed
      histogram) and the combined substructure fields (S1, S2 and S3;
      solid histogram), uncorrected for foreground contamination, and
      scaled to have unity area under the curves. These were
      derived by bilinear interpolation of 12Gyr isochrones for
      candidate RGB stars with $i_o \le 23.5$ (see text for
      details). Lower panel: Corrected MDFs for the substructure and
      disk fields (solid and dashed histograms, respectively), after
      subtraction of the ``halo'' MDF, suitably scaled. Dotted and
      dot-dashed vertical lines correspond to mean and median
      metallicities (respectively) for each MDF.}
  \end{center}
\end{figure*}

\clearpage
\newpage

\begin{figure*}
  \begin{center}
    \includegraphics[angle=270, width=12cm]{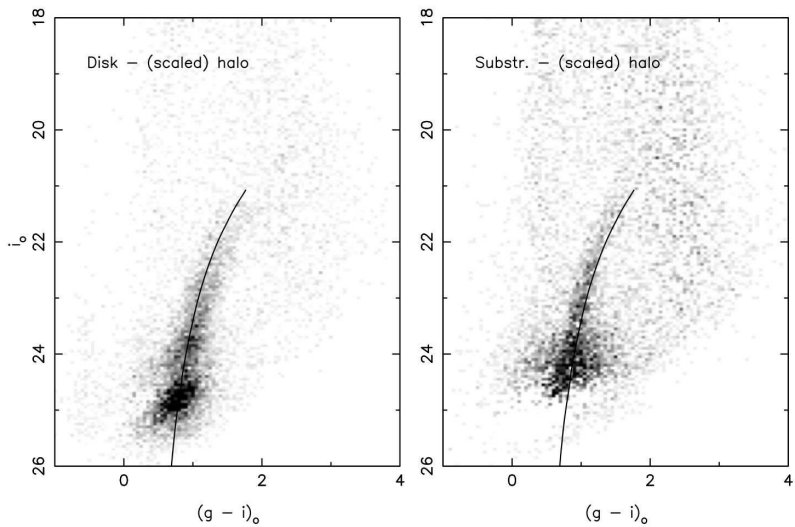}
    \caption{Background-corrected Hess diagrams for the disk (left
      panel) and combined substructure (right panel) fields. Here, the
      combined ``halo'' fields have been subtracted from the disk and
      substructure Hess diagrams using the same scaling as for the
      MDFs in the lower panel of Figure~10. Pixels are $0.05 \times
      0.05$ mags, and for clarity only positive residuals are shown on
      a linear scale. A 12Gyr, $[\alpha/Fe] = 0.0$, [Fe/H] = -1.2\,dex isochrones (the mean
      metallicity of the disk field) is overlaid in both panels for
      reference.}
  \end{center}
\end{figure*}

\clearpage
\newpage

\begin{figure*}
  \begin{center}
    \includegraphics[angle=270, width=12cm]{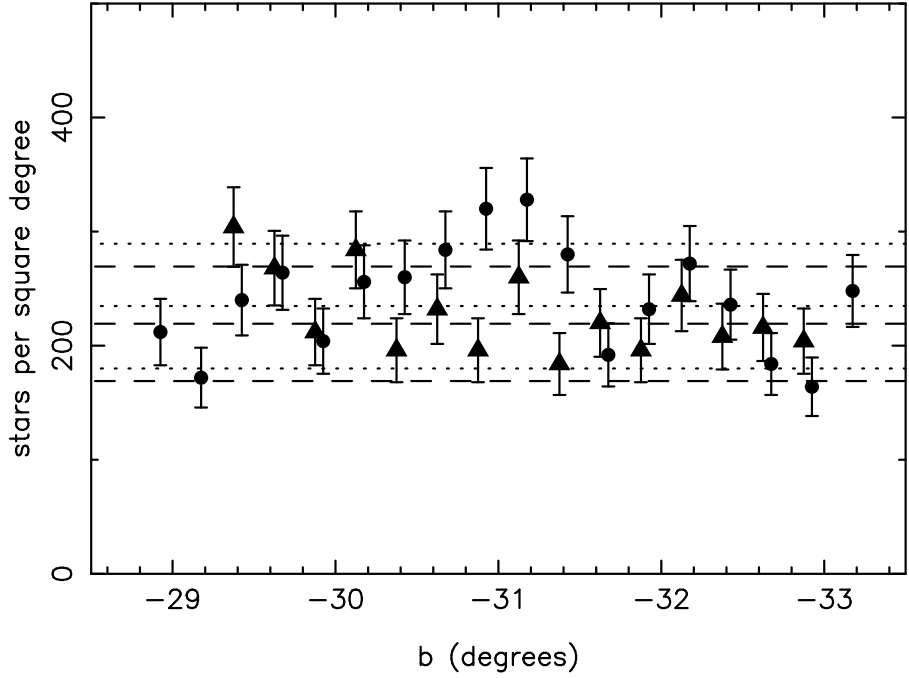}
    \caption{Number density of stars as a function of galactic
    latitude ($b$) in strip A (circles) and strip B (triangles), in
    the RGB photometric metallicity interval $-2.0 \le$ [Fe/H] $\le
    -1.0$\,dex. Most of these stars are expected to be Galactic
    foreground stars, and not genuine RGB stars in M33. Dotted and
    dashed lines show the mean and $1\sigma$ spreads for strips A and
    B, respectively. In this metallicity interval, the foreground
    contamination does not vary significantly between strips or with
    latitude.}
  \end{center}
\end{figure*}

\clearpage
\newpage

\begin{figure*}
  \begin{center}
    \includegraphics[angle=270, width=12cm]{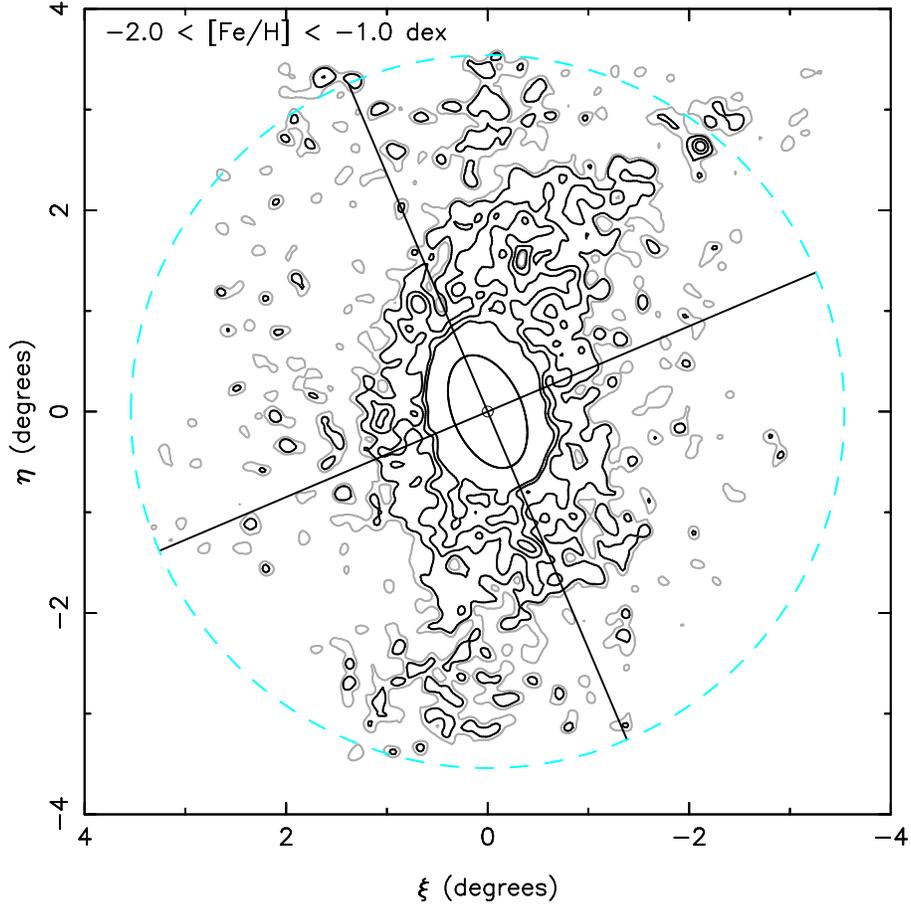}
    \caption{The density distribution of candidate RGB stars in the
    metallicity interval $-2.0 \le$ [Fe/H] $\le -1.0$\,dex, using an
    identical procedure to Figure~8. This metallicity cut was selected
    to optimally identify the M33 substructure. Black contours are 2,
    5, 8 and 12$\sigma$ above the background, corresponding to
    estimated surface brightness limits of $\mu_V = 32.5, 31.7, 31.2$
    and $30.6$\,mag\,arcsec$^{-2}$, respectively. The gray contour
    is $1\sigma$ above the background ($\mu_V \simeq
    33.0$\,mag\,arcsec$^{-2}$).}
  \end{center}
\end{figure*}

\clearpage
\newpage

\begin{figure*}
  \begin{center}
    \includegraphics[angle=270, width=12cm]{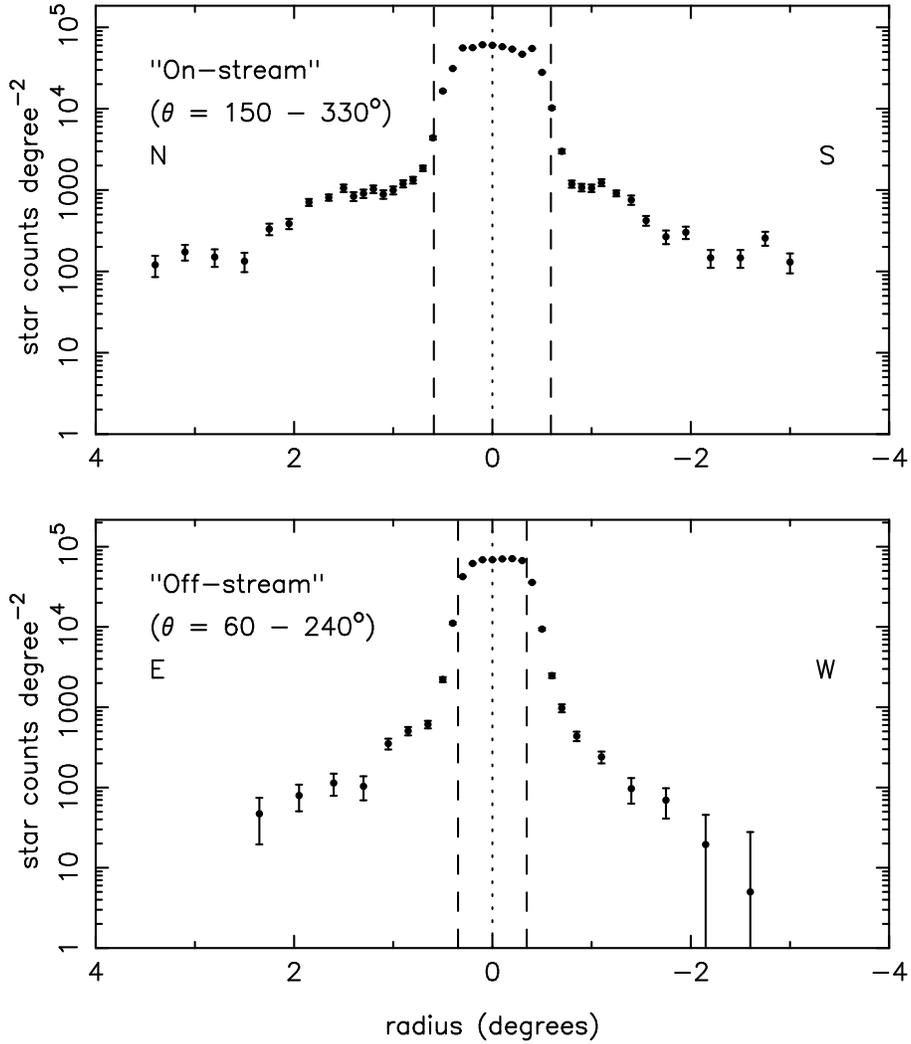}
    \caption{The radial profile of M33, constructed in 1 degree wide
      strips, in two orthogonal directions, using identical cuts as
      for Figure~13. The top panel shows the radial profile in an
      ``on-stream'' region, and the bottom panel shows the profile in
      an ``off-stream'' region. Here, $\theta$ refers to an angle,
      east-from-north, where $\theta = 0$ corresponds to the northern
      major axis of M33. Dashed lines correspond to the radius of the
      $\mu_B = 25$mag\,arcsec$^{-2}$ isophote. The extended structure
      is clearly visible in the top panel, and can also clearly be seen in the
      east in the lower panel. At larger radius in the lower panel,
      there are hints of a slowly declining component, presumably a
      M33 halo component.}
  \end{center}
\end{figure*}

\clearpage
\newpage

\begin{figure*}
  \begin{center}
    \includegraphics[angle=270, width=12cm]{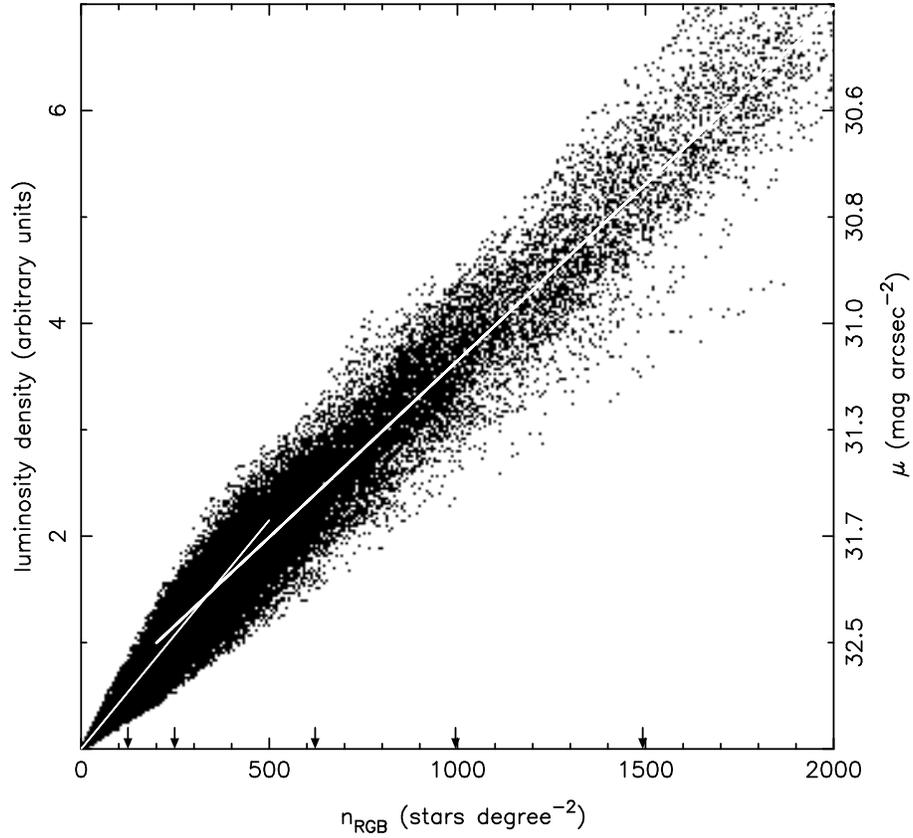}
    \caption{Relation between number density of RGB stars and
    luminosity density for Figure~13. The surface brightness scale on
    the right vertical axis is calculated using the transformation
    derived from Andromeda~I. Arrows indicate values of 1, 2,
    5, 8, and $12\sigma$ contours shown in Figure~13. See text for
    details.}
  \end{center}
\end{figure*}

\clearpage
\newpage

\begin{figure*}
  \begin{center}
    \includegraphics[angle=270, width=12cm]{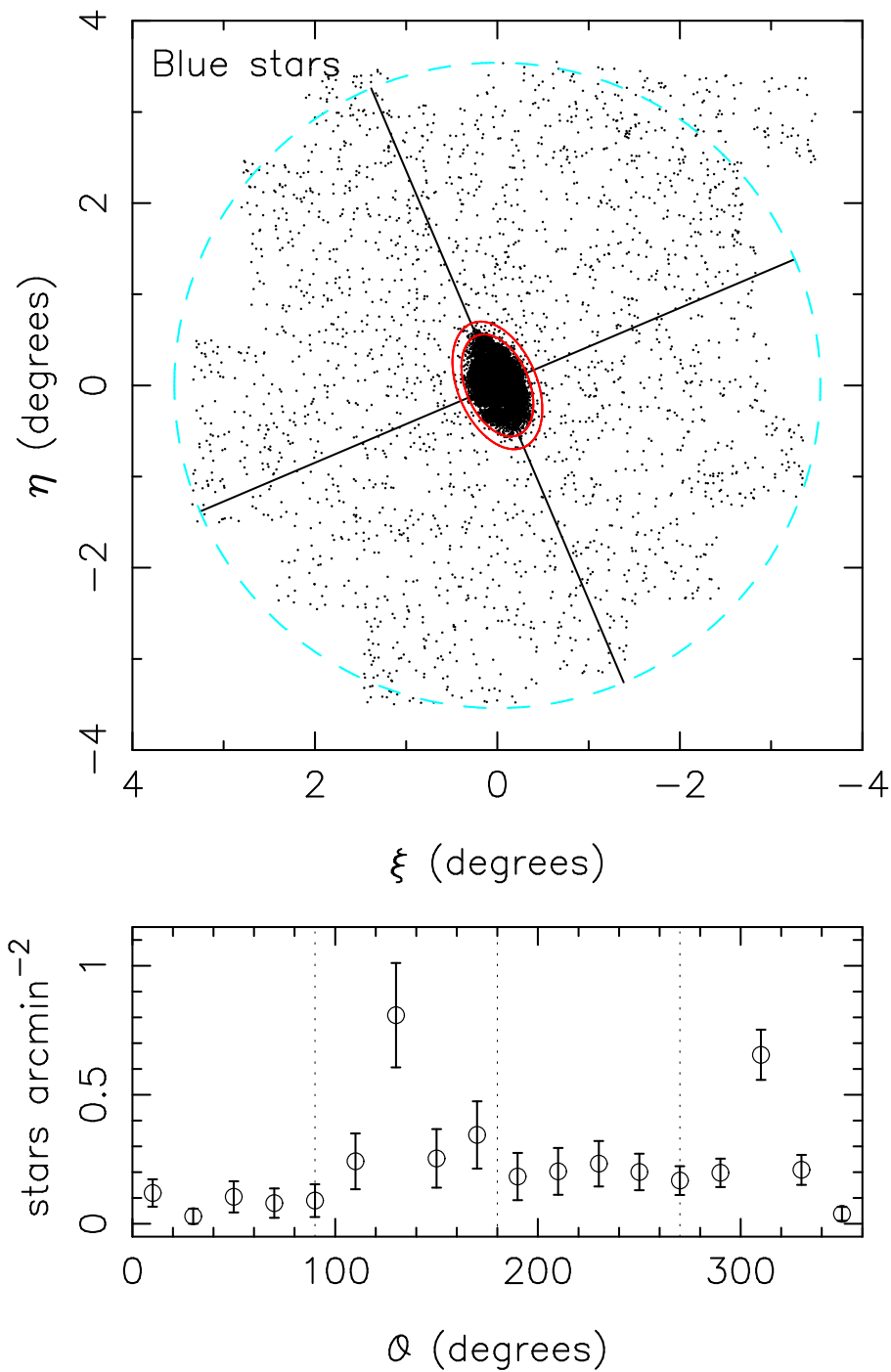}
    \caption{Upper panel: Tangent plane projection of the location of
    young, blue stars in M33, defined by $(g - i)_o < 0.0$ and $i_o <
    23.5$. An annulus in the outer disk is highlighted. Lower panel:
    Circular points show the density distribution of young stars in
    the annulus highlighted in the upper panel as a function of
    azimuth, where $\theta=0$ corresponds to the northern semi-major
    axis and $\theta=90^\circ$ corresponds to the eastern semi-minor
    axis.}
  \end{center}
\end{figure*}

\clearpage
\newpage

\begin{figure*}
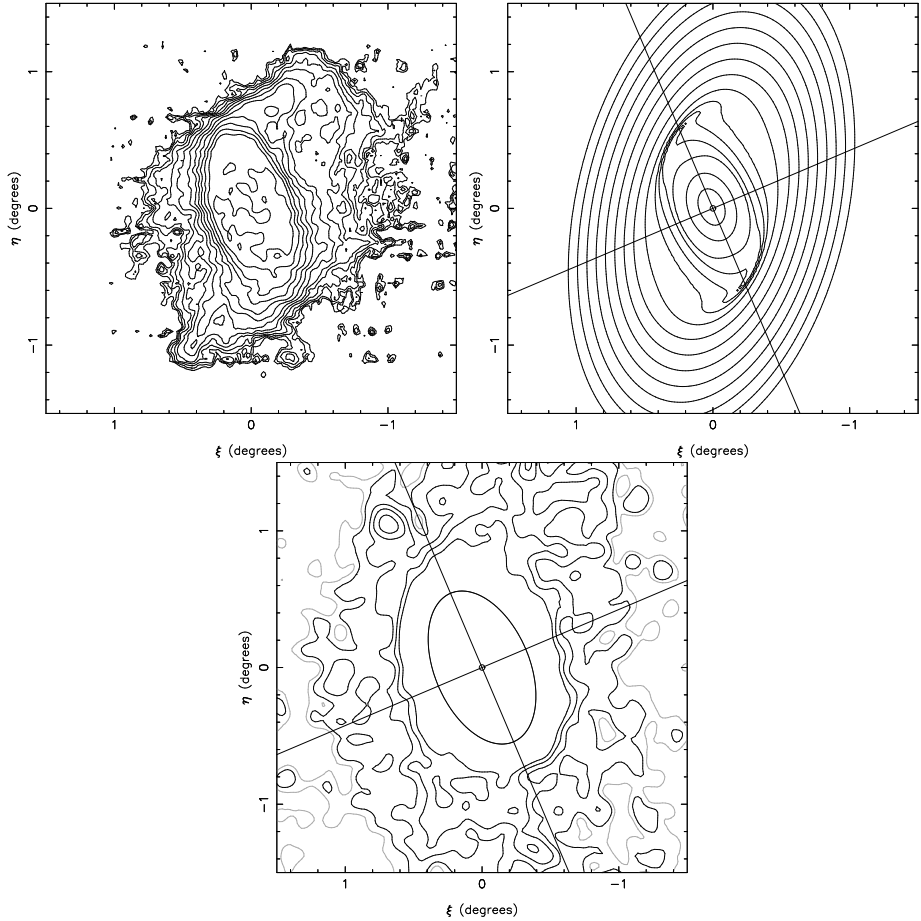

  \begin{center}
    \includegraphics[angle=270, width=6cm]{figure17a.ps}
    \includegraphics[angle=270, width=6cm]{figure17b.ps}
    \includegraphics[angle=270, width=6cm]{figure17c.ps}
    \caption{Left panel: M33 HI integrated column density, reproduced
    from Figure~1 of Putman et al. (2009). Contour levels are $8.3
    \times 1.5^n \times 10^{18}$ cm$^{-2}$ with $n = 0...13$. The
    lowest contour level is a $5\sigma$ detection threshold to a
    25\,km\,s$^{-1}$ feature. Middle panel: Simple model of the M33
    disk using the ring model of \cite{corbelli1997} for position and
    inclination as a function of $r$. Contour levels correspond to
    1\,magnitude intervals in surface brightness. Right panel: The M33
    stellar substructure surface brightness map. Contours are the same
    as Figure~13; the highest surface brightness contour has $\mu_V
    \simeq 30.7$\,mags\,arcsec$^{-2}$ (Table~1).}
  \end{center}
\end{figure*}

\clearpage
\newpage

\begin{table}
\begin{center}
\begin{tabular*}{0.6\textwidth}{ccc}
\hline
    & \multicolumn{2}{c}{$\mu_V$ mag\,arcsec$^{-2}$} \\
Level ($\sigma$) & And~I scale & And~III scale \\
\hline
1 & 32.8 (33.2) & 32.9 (33.3)\\
2 & 32.3 (32.4) & 32.5 (32.5)\\
5 & 31.6 (31.4) & 31.7 (31.6)\\
8 & 31.1 (30.9) & 31.2 (31.0)\\
12& 30.7 (30.5) & 30.8 (30.6)\\
\hline
\end{tabular*}
\caption{Conversions between contour levels in the stellar density map
(Figure~13) to surface brightness thresholds for different
empirical transformations. See text for details.}
\end{center}
\end{table}


\clearpage
\newpage
\begin{thebibliography}{59}
\expandafter\ifx\csname natexlab\endcsname\relax\def\natexlab#1{#1}\fi

\bibitem[{{Barker} {et~al.}(2007){Barker}, {Sarajedini}, {Geisler}, {Harding},
  \& {Schommer}}]{barker2007}
{Barker}, M.~K., {Sarajedini}, A., {Geisler}, D., {Harding}, P., \& {Schommer},
  R. 2007, \aj, 133, 1125

\bibitem[{{Bohlin}(2007)}]{bohlin2007}
{Bohlin}, R.~C. 2007, in Astronomical Society of the Pacific Conference Series,
  Vol. 364, The Future of Photometric, Spectrophotometric and Polarimetric
  Standardization, ed. C.~{Sterken}, 315

\bibitem[{{Bonanos} {et~al.}(2006){Bonanos}, {Stanek}, {Kudritzki}, {Macri},
  {Sasselov}, {Kaluzny}, {Stetson}, {Bersier}, {Bresolin}, {Matheson},
  {Mochejska}, {Przybilla}, {Szentgyorgyi}, {Tonry}, \& {Torres}}]{bonanos2006}
{Bonanos}, A.~Z., {Stanek}, K.~Z., {Kudritzki}, R.~P., {Macri}, L.~M.,
  {Sasselov}, D.~D., {Kaluzny}, J., {Stetson}, P.~B., {Bersier}, D.,
  {Bresolin}, F., {Matheson}, T., {Mochejska}, B.~J., {Przybilla}, N.,
  {Szentgyorgyi}, A.~H., {Tonry}, J., \& {Torres}, G. 2006, \apj, 652, 313

\bibitem[{{Bothun}(1992)}]{bothun1992}
{Bothun}, G.~D. 1992, \aj, 103, 104

\bibitem[{{Brunthaler} {et~al.}(2005){Brunthaler}, {Reid}, {Falcke},
  {Greenhill}, \& {Henkel}}]{brunthaler2005}
{Brunthaler}, A., {Reid}, M.~J., {Falcke}, H., {Greenhill}, L.~J., \& {Henkel},
  C. 2005, Science, 307, 1440

\bibitem[{{Chandar} {et~al.}(2002){Chandar}, {Bianchi}, {Ford}, \&
  {Sarajedini}}]{chandar2002}
{Chandar}, R., {Bianchi}, L., {Ford}, H.~C., \& {Sarajedini}, A. 2002, \apj,
  564, 712

\bibitem[{{Chapman} {et~al.}(2006){Chapman}, {Ibata}, {Lewis}, {Ferguson},
  {Irwin}, {McConnachie}, \& {Tanvir}}]{chapman2006}
{Chapman}, S.~C., {Ibata}, R., {Lewis}, G.~F., {Ferguson}, A.~M.~N., {Irwin},
  M., {McConnachie}, A., \& {Tanvir}, N. 2006, \apj, 653, 255

\bibitem[{{Chiba} \& {Beers}(2000)}]{chiba2000}
{Chiba}, M. \& {Beers}, T.~C. 2000, \aj, 119, 2843

\bibitem[{{Ciardullo} {et~al.}(2004){Ciardullo}, {Durrell}, {Laychak},
  {Herrmann}, {Moody}, {Jacoby}, \& {Feldmeier}}]{ciardullo2004}
{Ciardullo}, R., {Durrell}, P.~R., {Laychak}, M.~B., {Herrmann}, K.~A.,
  {Moody}, K., {Jacoby}, G.~H., \& {Feldmeier}, J.~J. 2004, \apj, 614, 167

\bibitem[{{Cioni}(2009)}]{cioni2009}
{Cioni}, M. 2009, \aap, 506, 1137

\bibitem[Clem et al.(2008)]{clem2008} Clem, J.~L., Vanden Berg, 
D.~A., \& Stetson, P.~B.\ 2008, \aj, 135, 682 

\bibitem[{{Corbelli}(2003)}]{corbelli2003}
{Corbelli}, E. 2003, \mnras, 342, 199

\bibitem[{{Corbelli} \& {Salucci}(2000)}]{corbelli2000}
{Corbelli}, E. \& {Salucci}, P. 2000, \mnras, 311, 441

\bibitem[{{Corbelli} \& {Schneider}(1997)}]{corbelli1997}
{Corbelli}, E. \& {Schneider}, S.~E. 1997, \apj, 479, 244

\bibitem[{{Da Costa} {et~al.}(2002){Da Costa}, {Armandroff}, \&
  {Caldwell}}]{dacosta2002}
{Da Costa}, G.~S., {Armandroff}, T.~E., \& {Caldwell}, N. 2002, \aj, 124, 332

\bibitem[{{Da Costa} {et~al.}(1996){Da Costa}, {Armandroff}, {Caldwell}, \&
  {Seitzer}}]{dacosta1996}
{Da Costa}, G.~S., {Armandroff}, T.~E., {Caldwell}, N., \& {Seitzer}, P. 1996,
  \aj, 112, 2576

\bibitem[{{Dotter} {et~al.}(2007){Dotter}, {Chaboyer}, {Jevremovi{\'c}},
  {Baron}, {Ferguson}, {Sarajedini}, \& {Anderson}}]{dotter2007}
{Dotter}, A., {Chaboyer}, B., {Jevremovi{\'c}}, D., {Baron}, E., {Ferguson},
  J.~W., {Sarajedini}, A., \& {Anderson}, J. 2007, \aj, 134, 376

\bibitem[{{Dotter} {et~al.}(2008){Dotter}, {Chaboyer}, {Jevremovi{\'c}},
  {Kostov}, {Baron}, \& {Ferguson}}]{dotter2008}
{Dotter}, A., {Chaboyer}, B., {Jevremovi{\'c}}, D., {Kostov}, V., {Baron}, E.,
  \& {Ferguson}, J.~W. 2008, \apjs, 178, 89

\bibitem[{{Ferguson} {et~al.}(2007){Ferguson}, {Irwin}, {Chapman}, {Ibata},
  {Lewis}, \& {Tanvir}}]{ferguson2007}
{Ferguson}, A., {Irwin}, M., {Chapman}, S., {Ibata}, R., {Lewis}, G., \&
  {Tanvir}, N. 2007, {Resolving the Stellar Outskirts of M31 and M33} (ISLAND
  UNIVERSES, Astrophysics and Space Science Proceedings.~ ISBN
  978-1-4020-5572-0.~Springer, 2007, p.~239), 239

\bibitem[{{Ferguson} {et~al.}(2002){Ferguson}, {Irwin}, {Ibata}, {Lewis}, \&
  {Tanvir}}]{ferguson2002}
{Ferguson}, A.~M.~N., {Irwin}, M.~J., {Ibata}, R.~A., {Lewis}, G.~F., \&
  {Tanvir}, N.~R. 2002, \aj, 124, 1452

\bibitem[{{Galleti} {et~al.}(2004){Galleti}, {Bellazzini}, \&
  {Ferraro}}]{galleti2004}
{Galleti}, S., {Bellazzini}, M., \& {Ferraro}, F.~R. 2004, \aap, 423, 925

\bibitem[Girardi et 
al.(2004)]{girardi2004} Girardi, L., Grebel, E.~K., Odenkirchen, M., \& Chiosi, C.\ 2004, \aap, 422, 205 

\bibitem[{{Huxor} {et~al.}(2009){Huxor}, {Ferguson}, {Barker}, {Tanvir},
  {Irwin}, {Chapman}, {Ibata}, \& {Lewis}}]{huxor2009}
{Huxor}, A., {Ferguson}, A.~M.~N., {Barker}, M.~K., {Tanvir}, N.~R., {Irwin},
  M.~J., {Chapman}, S.~C., {Ibata}, R., \& {Lewis}, G. 2009, \apjl, 698, L77

\bibitem[{{Ibata} {et~al.}(2001){Ibata}, {Irwin}, {Lewis}, {Ferguson}, \&
  {Tanvir}}]{ibata2001a}
{Ibata}, R., {Irwin}, M., {Lewis}, G., {Ferguson}, A.~M.~N., \& {Tanvir}, N.
  2001, \nat, 412, 49

\bibitem[{{Ibata} {et~al.}(2007){Ibata}, {Martin}, {Irwin}, {Chapman},
  {Ferguson}, {Lewis}, \& {McConnachie}}]{ibata2007}
{Ibata}, R., {Martin}, N.~F., {Irwin}, M., {Chapman}, S., {Ferguson}, A.~M.~N.,
  {Lewis}, G.~F., \& {McConnachie}, A.~W. 2007, \apj, 671, 1591

\bibitem[{{Irwin} \& {Lewis}(2001)}]{irwin2001}
{Irwin}, M. \& {Lewis}, J. 2001, New Astronomy Review, 45, 105

\bibitem[{{Irwin} {et~al.}(2008){Irwin}, {Ferguson}, {Huxor}, {Tanvir},
  {Ibata}, \& {Lewis}}]{irwin2008}
{Irwin}, M.~J., {Ferguson}, A.~M.~N., {Huxor}, A.~P., {Tanvir}, N.~R., {Ibata},
  R.~A., \& {Lewis}, G.~F. 2008, \apjl, 676, L17

\bibitem[{{Irwin} {et~al.}(2005){Irwin}, {Ferguson}, {Ibata}, {Lewis}, \&
  {Tanvir}}]{irwin2005}
{Irwin}, M.~J., {Ferguson}, A.~M.~N., {Ibata}, R.~A., {Lewis}, G.~F., \&
  {Tanvir}, N.~R. 2005, \apjl, 628, L105

\bibitem[{{Kalirai} {et~al.}(2006){Kalirai}, {Gilbert}, {Guhathakurta},
  {Majewski}, {Ostheimer}, {Rich}, {Cooper}, {Reitzel}, \&
  {Patterson}}]{kalirai2006b}
{Kalirai}, J.~S., {Gilbert}, K.~M., {Guhathakurta}, P., {Majewski}, S.~R.,
  {Ostheimer}, J.~C., {Rich}, R.~M., {Cooper}, M.~C., {Reitzel}, D.~B., \&
  {Patterson}, R.~J. 2006, \apj, 648, 389

\bibitem[{{Magrini} {et~al.}(2009){Magrini}, {Stanghellini}, \&
  {Villaver}}]{magrini2009}
{Magrini}, L., {Stanghellini}, L., \& {Villaver}, E. 2009, \apj, 696, 729

\bibitem[{{Martin} {et~al.}(2007){Martin}, {Ibata}, \& {Irwin}}]{martin2007}
{Martin}, N.~F., {Ibata}, R.~A., \& {Irwin}, M. 2007, \apjl, 668, L123

\bibitem[{{Martin} {et~al.}(2006){Martin}, {Ibata}, {Irwin}, {Chapman},
  {Lewis}, {Ferguson}, {Tanvir}, \& {McConnachie}}]{martin2006}
{Martin}, N.~F., {Ibata}, R.~A., {Irwin}, M.~J., {Chapman}, S., {Lewis}, G.~F.,
  {Ferguson}, A.~M.~N., {Tanvir}, N., \& {McConnachie}, A.~W. 2006, \mnras,
  371, 1983

\bibitem[{{Martin} {et~al.}(2009){Martin}, {McConnachie}, {Irwin}, {Widrow},
  {Ferguson}, {Ibata}, {Dubinski}, {Babul}, {Chapman}, {Fardal}, {Lewis},
  {Navarro}, \& {Rich}}]{martin2009}
{Martin}, N.~F., {McConnachie}, A.~W., {Irwin}, M., {Widrow}, L.~M.,
  {Ferguson}, A.~M.~N., {Ibata}, R.~A., {Dubinski}, J., {Babul}, A., {Chapman},
  S., {Fardal}, M., {Lewis}, G.~F., {Navarro}, J., \& {Rich}, R.~M. 2009, \apj,
  705, 758

\bibitem[{{Mateo}(1998)}]{mateo1998a}
{Mateo}, M.~L. 1998, \araa, 36, 435

\bibitem[{{McConnachie} {et~al.}(2007){McConnachie}, {Arimoto}, \&
  {Irwin}}]{mcconnachie2007a}
{McConnachie}, A.~W., {Arimoto}, N., \& {Irwin}, M. 2007, \mnras, 379, 379

\bibitem[{{McConnachie} {et~al.}(2006){McConnachie}, {Chapman}, {Ibata},
  {Ferguson}, {Irwin}, {Lewis}, {Tanvir}, \& {Martin}}]{mcconnachie2006c}
{McConnachie}, A.~W., {Chapman}, S.~C., {Ibata}, R.~A., {Ferguson}, A.~M.~N.,
  {Irwin}, M.~J., {Lewis}, G.~F., {Tanvir}, N.~R., \& {Martin}, N. 2006, \apjl,
  647, L25

\bibitem[{{McConnachie} {et~al.}(2008){McConnachie}, {Huxor}, {Martin},
  {Irwin}, {Chapman}, {Fahlman}, {Ferguson}, {Ibata}, {Lewis}, {Richer}, \&
  {Tanvir}}]{mcconnachie2008b}
{McConnachie}, A.~W., {Huxor}, A., {Martin}, N.~F., {Irwin}, M.~J., {Chapman},
  S.~C., {Fahlman}, G., {Ferguson}, A.~M.~N., {Ibata}, R.~A., {Lewis}, G.~F.,
  {Richer}, H., \& {Tanvir}, N.~R. 2008, \apj, 688, 1009

\bibitem[{{McConnachie} \& {Irwin}(2006)}]{mcconnachie2006a}
{McConnachie}, A.~W. \& {Irwin}, M.~J. 2006, \mnras, 365, 902

\bibitem[{{McConnachie} {et~al.}(2004{\natexlab{a}}){McConnachie}, {Irwin},
  {Ferguson}, {Ibata}, {Lewis}, \& {Tanvir}}]{mcconnachie2004a}
{McConnachie}, A.~W., {Irwin}, M.~J., {Ferguson}, A.~M.~N., {Ibata}, R.~A.,
  {Lewis}, G.~F., \& {Tanvir}, N. 2004{\natexlab{a}}, \mnras, 350, 243

\bibitem[{{McConnachie} {et~al.}(2005){McConnachie}, {Irwin}, {Ferguson},
  {Ibata}, {Lewis}, \& {Tanvir}}]{mcconnachie2005a}
---. 2005, \mnras, 356, 979

\bibitem[{{McConnachie} {et~al.}(2009){McConnachie}, {Irwin}, {Ibata},
  {Dubinski}, {Widrow}, {Martin}, {C{\^o}t{\'e}}, {Dotter}, {Navarro},
  {Ferguson}, {Puzia}, {Lewis}, {Babul}, {Barmby}, {Bienaym{\'e}}, {Chapman},
  {Cockcroft}, {Collins}, {Fardal}, {Harris}, {Huxor}, {Mackey},
  {Pe{\~n}arrubia}, {Rich}, {Richer}, {Siebert}, {Tanvir}, {Valls-Gabaud}, \&
  {Venn}}]{mcconnachie2009b}
{McConnachie}, A.~W., {Irwin}, M.~J., {Ibata}, R.~A., {Dubinski}, J., {Widrow},
  L.~M., {Martin}, N.~F., {C{\^o}t{\'e}}, P., {Dotter}, A.~L., {Navarro},
  J.~F., {Ferguson}, A.~M.~N., {Puzia}, T.~H., {Lewis}, G.~F., {Babul}, A.,
  {Barmby}, P., {Bienaym{\'e}}, O., {Chapman}, S.~C., {Cockcroft}, R.,
  {Collins}, M.~L.~M., {Fardal}, M.~A., {Harris}, W.~E., {Huxor}, A., {Mackey},
  A.~D., {Pe{\~n}arrubia}, J., {Rich}, R.~M., {Richer}, H.~B., {Siebert}, A.,
  {Tanvir}, N., {Valls-Gabaud}, D., \& {Venn}, K.~A. 2009, \nat, 461, 66

\bibitem[{{McConnachie} {et~al.}(2003){McConnachie}, {Irwin}, {Ibata},
  {Ferguson}, {Lewis}, \& {Tanvir}}]{mcconnachie2003}
{McConnachie}, A.~W., {Irwin}, M.~J., {Ibata}, R.~A., {Ferguson}, A.~M.~N.,
  {Lewis}, G.~F., \& {Tanvir}, N. 2003, \mnras, 343, 1335

\bibitem[{{McConnachie} {et~al.}(2004{\natexlab{b}}){McConnachie}, {Irwin},
  {Lewis}, {Ibata}, {Chapman}, {Ferguson}, \& {Tanvir}}]{mcconnachie2004b}
{McConnachie}, A.~W., {Irwin}, M.~J., {Lewis}, G.~F., {Ibata}, R.~A.,
  {Chapman}, S.~C., {Ferguson}, A.~M.~N., \& {Tanvir}, N.~R.
  2004{\natexlab{b}}, \mnras, 351, L94

\bibitem[{{McLean} \& {Liu}(1996)}]{mclean1996}
{McLean}, I.~S. \& {Liu}, T. 1996, \apj, 456, 499

\bibitem[{{Minniti} {et~al.}(1993){Minniti}, {Olszewski}, \&
  {Rieke}}]{minniti1993}
{Minniti}, D., {Olszewski}, E.~W., \& {Rieke}, M. 1993, \apjl, 410, L79

\bibitem[{{Purcell} {et~al.}(2007){Purcell}, {Bullock}, \&
  {Zentner}}]{purcell2007}
{Purcell}, C.~W., {Bullock}, J.~S., \& {Zentner}, A.~R. 2007, \apj, 666, 20

\bibitem[{{Putman} {et~al.}(2009){Putman}, {Peek}, {Muratov}, {Gnedin}, {Hsu},
  {Douglas}, {Heiles}, {Stanimirovic}, {Korpela}, \& {Gibson}}]{putman2009}
{Putman}, M.~E., {Peek}, J.~E.~G., {Muratov}, A., {Gnedin}, O.~Y., {Hsu}, W.,
  {Douglas}, K.~A., {Heiles}, C., {Stanimirovic}, S., {Korpela}, E.~J., \&
  {Gibson}, S.~J. 2009, \apj, 703, 1486

\bibitem[{{Regan} \& {Vogel}(1994)}]{regan1994}
{Regan}, M.~W. \& {Vogel}, S.~N. 1994, \apj, 434, 536

\bibitem[{{Regnault} {et~al.}(2009){Regnault}, {Conley}, {Guy}, {Sullivan},
  {Cuillandre}, {Astier}, {Balland}, {Basa}, {Carlberg}, {Fouchez}, {Hardin},
  {Hook}, {Howell}, {Pain}, {Perrett}, \& {Pritchet}}]{regnault2009}
{Regnault}, N., {Conley}, A., {Guy}, J., {Sullivan}, M., {Cuillandre}, J.,
  {Astier}, P., {Balland}, C., {Basa}, S., {Carlberg}, R.~G., {Fouchez}, D.,
  {Hardin}, D., {Hook}, I.~M., {Howell}, D.~A., {Pain}, R., {Perrett}, K., \&
  {Pritchet}, C.~J. 2009, \aap, 506, 999

\bibitem[{{Rejkuba} {et~al.}(2006){Rejkuba}, {da Costa}, {Jerjen}, {Zoccali},
  \& {Binggeli}}]{rejkuba2006}
{Rejkuba}, M., {da Costa}, G.~S., {Jerjen}, H., {Zoccali}, M., \& {Binggeli},
  B. 2006, \aap, 448, 983

\bibitem[{{Renzini} \& {Fusi Pecci}(1988)}]{renzini1988}
{Renzini}, A. \& {Fusi Pecci}, F. 1988, \araa, 26, 199

\bibitem[{{Rocha-Pinto} {et~al.}(2004){Rocha-Pinto}, {Majewski}, {Skrutskie},
  {Crane}, \& {Patterson}}]{rochapinto2004}
{Rocha-Pinto}, H.~J., {Majewski}, S.~R., {Skrutskie}, M.~F., {Crane}, J.~D., \&
  {Patterson}, R.~J. 2004, \apj, 615, 732

\bibitem[{{Rogstad} {et~al.}(1976){Rogstad}, {Wright}, \&
  {Lockhart}}]{rogstad1976}
{Rogstad}, D.~H., {Wright}, M.~C.~H., \& {Lockhart}, I.~A. 1976, \apj, 204, 703

\bibitem[{{Sarajedini} {et~al.}(2006){Sarajedini}, {Barker}, {Geisler},
  {Harding}, \& {Schommer}}]{sarajedini2006}
{Sarajedini}, A., {Barker}, M.~K., {Geisler}, D., {Harding}, P., \& {Schommer},
  R. 2006, \aj, 132, 1361

\bibitem[{{Schlegel} {et~al.}(1998){Schlegel}, {Finkbeiner}, \&
  {Davis}}]{schlegel1998}
{Schlegel}, D.~J., {Finkbeiner}, D.~P., \& {Davis}, M. 1998, \apj, 500, 525

\bibitem[{{Stonkut{\.e}} {et~al.}(2008){Stonkut{\.e}}, {Vansevi{\v c}ius},
  {Arimoto}, {Hasegawa}, {Narbutis}, {Tamura}, {Jablonka}, {Ohta}, \&
  {Yamada}}]{stonkute2008}
{Stonkut{\.e}}, R., {Vansevi{\v c}ius}, V., {Arimoto}, N., {Hasegawa}, T.,
  {Narbutis}, D., {Tamura}, N., {Jablonka}, P., {Ohta}, K., \& {Yamada}, Y.
  2008, \aj, 135, 1482

\bibitem[{{Tiede} {et~al.}(2004){Tiede}, {Sarajedini}, \& {Barker}}]{tiede2004}
{Tiede}, G.~P., {Sarajedini}, A., \& {Barker}, M.~K. 2004, \aj, 128, 224

\bibitem[{{U} {et~al.}(2009){U}, {Urbaneja}, {Kudritzki}, {Jacobs}, {Bresolin},
  \& {Przybilla}}]{u2009}
{U}, V., {Urbaneja}, M.~A., {Kudritzki}, R., {Jacobs}, B.~A., {Bresolin}, F.,
  \& {Przybilla}, N. 2009, \apj, 704, 1120

\bibitem[{{Zucker} {et~al.}(2004{\natexlab{a}}){Zucker}, {Kniazev}, {Bell},
  {Mart{\'{\i}}nez-Delgado}, {Grebel}, {Rix}, {Rockosi}, {Holtzman},
  {Walterbos}, {Annis}, {York}, {Ivezi{\' c}}, {Brinkmann}, {Brewington},
  {Harvanek}, {Hennessy}, {Kleinman}, {Krzesinski}, {Long}, {Newman}, {Nitta},
  \& {Snedden}}]{zucker2004a}
{Zucker}, D.~B., {Kniazev}, A.~Y., {Bell}, E.~F., {Mart{\'{\i}}nez-Delgado},
  D., {Grebel}, E.~K., {Rix}, H., {Rockosi}, C.~M., {Holtzman}, J.~A.,
  {Walterbos}, R.~A.~M., {Annis}, J., {York}, D.~G., {Ivezi{\' c}}, {\v Z}.,
  {Brinkmann}, J., {Brewington}, H., {Harvanek}, M., {Hennessy}, G.,
  {Kleinman}, S.~J., {Krzesinski}, J., {Long}, D., {Newman}, P.~R., {Nitta},
  A., \& {Snedden}, S.~A. 2004{\natexlab{a}}, \apjl, 612, L121

\bibitem[{{Zucker} {et~al.}(2004{\natexlab{b}}){Zucker}, {Kniazev}, {Bell},
  {Mart{\'{\i}}nez-Delgado}, {Grebel}, {Rix}, {Rockosi}, {Holtzman},
  {Walterbos}, {Ivezi{\' c}}, {Brinkmann}, {Brewington}, {Harvanek},
  {Kleinman}, {Krzesinski}, {Lamb}, {Long}, {Newman}, {Nitta}, \&
  {Snedden}}]{zucker2004b}
{Zucker}, D.~B., {Kniazev}, A.~Y., {Bell}, E.~F., {Mart{\'{\i}}nez-Delgado},
  D., {Grebel}, E.~K., {Rix}, H., {Rockosi}, C.~M., {Holtzman}, J.~A.,
  {Walterbos}, R.~A.~M., {Ivezi{\' c}}, {\v Z}., {Brinkmann}, J., {Brewington},
  H., {Harvanek}, M., {Kleinman}, S.~J., {Krzesinski}, J., {Lamb}, D.~Q.,
  {Long}, D., {Newman}, P.~R., {Nitta}, A., \& {Snedden}, S.~A.
  2004{\natexlab{b}}, \apjl, 612, L117

\end{thebibliography}
\end{document}